\documentclass[11pt,twoside,a4paper]{article}
\usepackage{authblk}
\usepackage{amssymb,amsmath}
\numberwithin{figure}{subsection}
\numberwithin{table}{subsection}
\numberwithin{equation}{subsection}

\usepackage{graphicx}
\usepackage[dvips]{epsfig}
\usepackage[dvipsnames]{xcolor}
\usepackage{amssymb}
\usepackage{rotating}
\usepackage{lscape}
\usepackage{a4wide}
\usepackage{mdframed}
\usepackage[super]{nth}
\usepackage{tikz}
\usepackage{siunitx}
\usepackage{subfigure}
\usepackage{vruler}
\usepackage[super]{nth}
\usepackage[sort&compress,numbers]{natbib}
\usepackage[colorlinks]{hyperref}
\hypersetup{
    colorlinks=true, 
    linktoc=all,     
    linkcolor=blue,  
    citecolor=magenta,
    urlcolor=blue	
}
\usepackage{lmodern}
\usepackage[utf8]{inputenc}
\usepackage[T1]{fontenc}
\usepackage{threeparttable}

\title{Spin Physics at COSY (2021 - 2024 and beyond)\\
{\large -- Pathfinder investigations toward an EDM storage ring and Spin-for-FAIR --}}

\vfill

\author[1]{R.\,Gebel}
\author[1]{V.\,Hejny}
\author[1]{A.\,Kacharava}
\author[1,2]{A.\,Lehrach}
\author[3]{P.\,Lenisa}
\author[1]{A.\,Nass}
\author[1,2]{J.\,Pretz}
\author[1]{F.\,Rathmann}
\author[1]{H.\,Ströher}

\date{\today}

\affil[1]{\small Institut für Kernphysik, Forschungszentrum Jülich, 52428 Jülich, Germany}
\affil[2]{\small III. Physikalisches Institut B, RWTH Aachen University, 52056 Aachen, Germany}
\affil[3]{\small University of Ferrara and INFN, 44100 Ferrara, Italy}
\begin{document}

\maketitle

\begin{abstract}
The unique global feature of COSY is its ability to accelerate, store and manipulate polarized proton and deuteron beams. In the recent past, these beams have been used primarily for precision measurements, in particular in connection with the study of charged particle EDMs (Electric Dipole Moment) in storage rings. The role of COSY as a R\&D facility and for initial (static and oscillating) EDM measurements can hardly be overestimated. Unfortunately, as a consequence of the strategic decisions of Forschungszentrum Jülich and the subsequent "TransFAIR" agreement between FZJ and GSI Darmstadt, it is currently planned to stop the operation of COSY by the end of 2024. The various groups working with polarized beams at COSY felt it important to collect information on essential measurements to be performed until the termination of machine operation. These experiments, briefly described in this document along with an estimate of the beam time required, serve as pathfinder investigations toward an EDM storage ring and Spin for FAIR.

\end{abstract}
\thispagestyle{empty}
\cleardoublepage
\cleardoublepage
\tableofcontents
\cleardoublepage

\section{Introduction}
The COoler SYnchrotron (COSY) at the Institute for Nuclear Physics (IKP) of Forschungs\-zentrum Jülich (FZJ) is a storage ring, capable of accelerating proton or deuteron beams up to momenta of about \SI{3.7}{GeV/c}. It is equipped with both electron cooling and stochastic cooling to provide high-quality beams to internal and external target stations.

COSY's worldwide unique feature is its capability to accelerate, store and manipulate polarized proton and deuteron beams for hadron physics experiments, e.g., at ANKE and PAX\,\cite{Kacharava:2005wz,pax:ERC,Wilkin2017} and, more recently, for precision measurements, in particular in connection with the research into charged particle EDMs (Electric Dipole Moment) in storage rings\,\cite{jedi:ERC,cpEDM2021}. The role of COSY as R\&D facility as well as for first (static and oscillating) EDM measurements can hardly be overestimated and has been acknowledged, e.g., in the update of the European Strategy for Particle Physics\,\cite{delib-europ-strat:2020}. It should also be noted that the milestones of the HGF Strategic Evaluation in early 2020 also comprise measurements with polarized beams at COSY\,\cite{HGF-eval:2020}.

Regrettably, as a result of the strategy decisions of Forschungszentrum Jülich and the subsequent „TransFAIR“ agreement between FZJ and GSI Darmstadt, the current plan is to stop COSY operations by the end of 2024. The different groups working with polarized beams at COSY thus felt it important to collect information about essential measurements which should be performed between now and the termination of the machine operation. 

In the following, these experiments are briefly described together with estimates of the required beam time for the information and consideration of the COSY Beam Advisory Committee (CBAC) and the Managements of FZJ and GSI.

%

A number of these proposals are related to the investigations in charged particle EDMs. During the last years the JEDI collaboration has performed a sequence of experiments on prolonging and understanding the spin coherence time (SCT) of cooled and bunched deuterons at a beam momentum of $p=970\,\mathrm{MeV/c}$. It has been shown that sextupole configurations leading to small chromaticities in the horizontal and vertical plane result in long SCTs in the order of $\tau=1000\,\mathrm{s}$\,\cite{PhysRevLett.117.054801, PhysRevAccelBeams.21.024201}. This work also resulted in a number of tools to measure the spin coherence time as well as the spin tune and to phase-lock the spin precession to an external RF resonator (RF solenoid, RF Wien filter)\,\cite{PhysRevLett.115.094801, PhysRevLett.119.014801}. In parallel, new technical equipment has been developed and installed in the COSY ring\,\cite{Fal-18-10,proposal-siberian-snake,Mueller:2020,Slim:2016pim,Slim:2020ufk}. Systematic studies on ring imperfections and spin dynamics\,\cite{Saleev:2017ecu,PhysRevAccelBeams.21.042002,proposal-Artem,PhysRevAccelBeams.23.024601}, beam-based alignment\,\cite{Wagner:2021}, and techniques that allow for the systematic evaluation of uncertainties\,\cite{Slim201752} completed these investigations. All the experimental work was accompanied by theoretical studies and lattice simulations of spin and particle tracking\,\cite{Rosenthal:2015,Poncza2019}. 

One major goal for the activities with deuterons at COSY is a first direct measurement of the deuteron electric dipole moment (EDM) using the waveguide RF Wien filter\,\cite{jedi2018_e005}. A first run took place in Q4 2018, a second one in Q1/Q2 2021. Data analysis is currently in progress. In addition, using the same experimental setup and similar techniques, a pilot run took place in Q2 2019 on the search for axion-like particles by means of on induced oscillating EDM.

The long term strategy for EDM searches utilizing storage rings is described in a feasibility study\,\cite{1812.08535, cpEDM2021} written as input for the discussion on the \textit{European Particle Physics Strategy Update 2018 – 2020}\,\cite{eustrategy,ESPP2019} and for the CERN \textit{Physics Beyond Colliders study group}\,\cite{Alemany2019}. It foresees a step-wise plan starting with the current COSY activities. The project continues with an intermediate prototype ring as a demonstrator for key technologies for an electric ring as well as for frozen spin, and has as a final goal a high precision all-electric storage ring for protons with counter-rotating beams. 


Consequently, the proposed JEDI experiments are either a continuation of existing efforts (axion search: Sec.\,\ref{sec:axion}, spin tune response to local steerer bumps: Sec.\,\ref{sec:spin-tune-orbit-bumps}) or preparations for future options (spin coherence time for protons: Sec.\,\ref{sec:spico-for-protons}, stochastic beam cooling: Sec.\,\ref{sec:cooling}) that can currently only be done at COSY. In addition, the perspectives for a first direct measurement of the proton electric dipole moment using the RF Wien filter technique is discussed (Sec.\,\ref{sec:protonedm}). 

Other proposals deal with a novel technique for the control of particle spins in storage rings called spin transparency (Sec.\,\ref{sec:spintransparency}), a highly flexible method to manipulate the spin direction in real time, and an experiment on longitudinal spin filtering in order to study the possibility of polarized antiprotons at FAIR (Sec.\,\ref{sec:spinfiltering}).

\cleardoublepage

\section{Spin physics proposals for COSY}
\label{sec:spin-physics-at-COSY}

\subsection{Spin-coherence time investigations for protons}
\label{sec:spico-for-protons}

\begin{center}
	V. Hejny$^1$, I. Keshelashvili$^1$, A. Lehrach$^1$, E.J. Stephenson$^2$, A. Wro\'nska$^3$\\
	\vspace{0.2cm}
	(for the JEDI collaboration)\\
	\vspace{0.2cm}
	$^1$ \textit{Institut für Kernphysik, Forschungszentrum Jülich, 52425 Jülich, Germany} \\
	$^2$ \textit{Center for Spacetime Symmetries, Indiana University, Bloomington, IN 47408, USA} \\
	$^3$ \textit{Marian Smoluchowski Institute of Physics, Jagiellonian University, 30-348 Krak\'ow, Poland} \\
\end{center}

\begin{mdframed}
	\begin{enumerate}
		\item \textbf{Estimated number of MD weeks: \textcolor{red}{6} }
		\item \textbf{Estimated number of BT weeks: \textcolor{red}{9} }
	\end{enumerate}
\end{mdframed}

A long spin coherence time is an essential prerequisite for any experiment which probes a polarization component perpendicular to the invariant spin axis. Consequently, a major part of the JEDI experimental program during the last years was on prolonging and understanding the spin coherence time (SCT) of cooled and bunched deuterons. It has been shown that long SCTs in the order of $\tau=1000\,\mathrm{s}$ can be achieved mainly by optimizing the sextupole settings, and, thus, the chromaticity\,\cite{PhysRevLett.117.054801, PhysRevAccelBeams.21.024201}. Before one approaches any EDM experiment on protons similar studies are mandatory as there are currently no data on proton spin coherence times. New experimental data are also urgently needed to benchmark the spin tracking codes for protons with respect to spin tune deviations and spin coherence time.

A number of things suggest that - compared to the deuteron - the proton is the more challenging case, including the larger anomalous moment and the larger spin-precession frequency, the greater abundance of intrinsic and imperfection spin resonances, and more complications involving the landscape of chromaticity that leads to shorter polarization lifetimes in the COSY ring. First results of more detailed simulation studies indicate that for the lattice configuration with minimized dispersion in the straights, which was used for the deuteron studies, a long spin coherence time could not be achieved by simply minimizing the chromaticities. Furthermore, the calculated required sextupole corrections for preserving the spin coherence lead to an unstable beam motion. 

In general, the spin coherence time (SCT) of an ensemble of particles is determined by the deviations of the spin precession rates as well as the spin precession axes of the individual particles with respect to the reference particle. These deviations have to be kept small in order to achieve a long SCT. Variations in the precession rate (or spin tune) are caused by the impact of path lengthening on the individual particles distributed in the transverse and longitudinal phase space. In addition, intrinsic spin resonances introduce a vertical phase-space dependent tilt of the rotation axis. This is described by the extension of the invariant spin vector (rotation axis of the reference particle) to the invariant spin field, which depends on the phase space position of a particle. If the distance to an intrinsic resonance is large (as for the deuteron case) and/or its strength is weak, these tilts become small. 

The strengths of intrinsic resonances strongly depend on the amplitudes of the vertical phase space motion. Thus, the spin motion of different particles is affected incoherently, which could introduce spin tune deviations. The impact of a single isolated intrinsic resonance on the spin tune has been discussed in detail in\,\cite{Rosenthal:2016zbf}. The simulations also predict a strong dependence of the SCT on the beam parameters like emittances and momentum spread paired with lattice parameters like chromaticities. It has also been shown that the resulting spin tune spread from intrinsic resonances can be compensated by an opposite deviation due to path lengthening by choosing a non-zero vertical chromaticity. 

\begin{figure}[ht]
	\begin{center}
		\includegraphics[width=0.48\textwidth,trim=15 0 0 0]{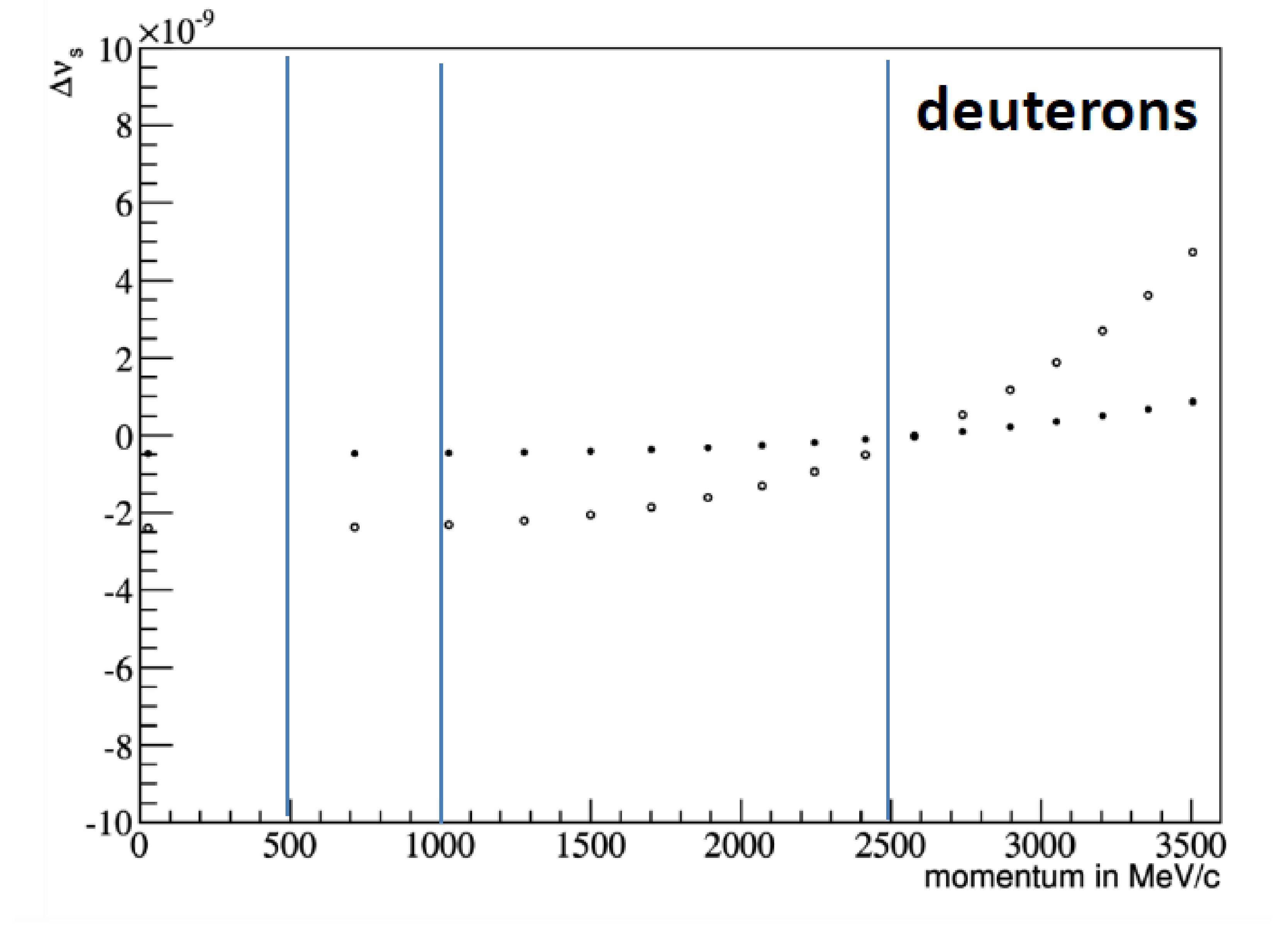}
		\includegraphics[width=0.48\textwidth]{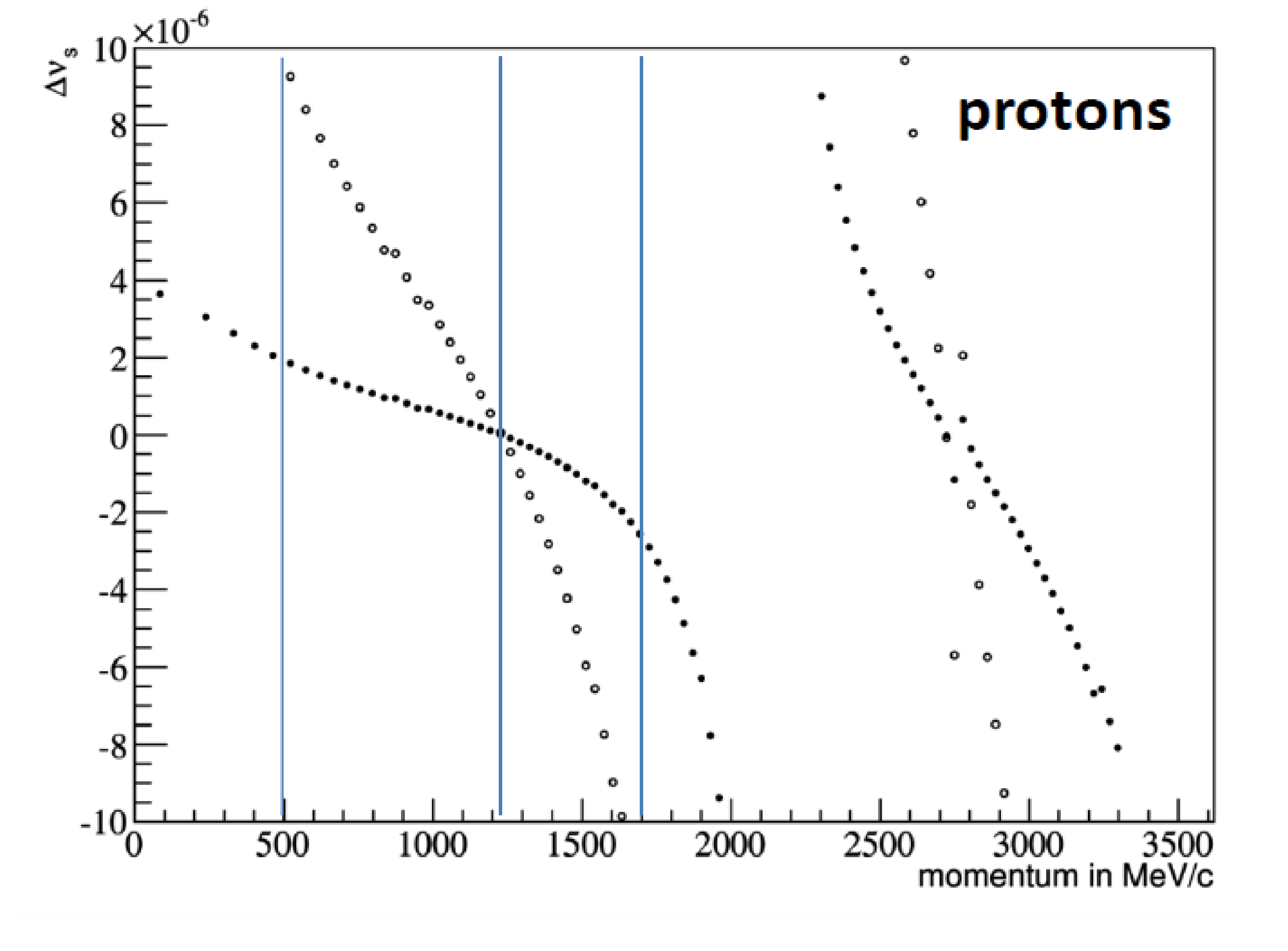}
		\caption{\label{Fig:SCT_PD} Spin tune deviations as function of beam momentum for a normalized emittance of 1~mm\,mrad (solid dots) and 5~mm\,mrad (open dots), respectively. The left plot shows deuterons, the right one protons. For each point an ensemble of 320 particle was randomly distributed in the six dimensional phase space. For protons the location of the 8- resonance is indicated. Note the different scale on the $y$-axis.}
	\end{center}
	
\end{figure}

The effect of intrinsic resonances on the SCT is much stronger for protons than for deuterons. 
Intrinsic resonances appear at $\gamma G = n \cdot P \pm (Q_y - 2)$, where $P$ is the super periodicity and $Q_y$ the vertical betatron tune. Detailed simulations over a wide momentum range have been performed to further investigate the contributions from the different intrinsic resonances \cite{Rosenthal:2016zbf}. The results are summarized in Fig.\,\ref{Fig:SCT_PD}. For comparison the deuteron case is shown in the left plot: there is no intrinsic resonance in the momentum range accessible with COSY and the spin tune deviations are in the order of $\nu_s \approx 10^{-9}$. For protons the situation is different: intrinsic resonances cause spin tune deviations that are larger by up to three orders of magnitude. The longest spin coherence times are expected to be reached at the zero crossings. The induced shifts of these zero crossing locations are proportional to the vertical chromaticity, which controls the amount of path lengthening for individual particles induced by intrinsic resonances. Adjusting the vertical chromaticity using sextupole magnets allows one to move the zero crossing location to a desired reference beam momentum. The actual value of the SCT finally depends on the beam emittances and the distance to the zero crossing momenta.

While it would be desirable to start at the location of such a zero crossing, also other constraints need to be considered.
\begin{itemize}
	\item As for the proposed investigations a normalized emittance of $\epsilon \leq 1\,\mathrm{mm\,mrad}$ is crucial, efficient beam cooling is mandatory. For all experiments with deuterons electron cooling was the method of choice and has been continuously optimized. Therefore, it should be also used for the initial run with protons, although this sets the upper limit of the beam energy to $T=140\,\mathrm{MeV}$. 
	\item Proton beam polarization measurements will be made using the JEPO polarimeter \cite{Mueller:2020} by recording forward elastic scattering events. The optimum operation range for protons is in a similar energy range as the electron cooling range and the average analyzing power within the acceptance 
	of the polarimeter drops with decreasing energy. 
\end{itemize}
This defines the (initial) proton kinetic energy as $T=140\,\mathrm{MeV}$. While this energy is still higher than the energies of the planned demonstrator ring and lower than the 232.8~MeV ($p=0.7007$~GeV/c) of the final proton EDM experiment \cite{1812.08535, cpEDM2021}, we expect that any results concerning polarization measurement and lifetime can be transferred to different energies and ring lattices once the simulation codes have been validated.

\subsection*{Beam Time Estimate}

The last COSY beam time with polarized protons was the pC database run \cite{jedi2018_e004.2} in summer 2018; however it was without cooling or a strong demand on beam intensity. We learned from the deuteron runs that it is advisable to switch off the e-cooler magnets after cooling to establish a better orbit, which needs a special setup. Also, a minimum number of particles of $N \geq 10^9$ is needed to perform a continuous measurement of the horizontal polarization. In the range up to $T=140\,\mathrm{MeV}$ intrinsic resonances are not present, however, higher order spin resonances as well as the imperfection resonance at $\nu_s = 2$ ($T=108\,\mathrm{MeV}$) have to be considered. Taking into account also the time for optimizing the polarized source, the experience from the deuteron runs tells us that this demands a prolonged machine development time of 2-3 weeks prior to each beam time.

In a first beam time and as an important first step all the techniques from the deuteron case have to be transferred to protons: extraction of the beam onto a carbon target by noise excitation to measure the beam polarization, spin rotation by means of an RF solenoid, and the continuous measurement of the spin tune and the degree of polarization of the spin-precessing particle ensemble. Subsequently, we can start investigating the effect of the COSY operation parameters (chromaticity, betatron tunes, cooling, lattice symmetries, etc.) on the SCT.

During the first run, the machine parameters will be chosen based on most recent simulations and the discussed boundary conditions on cooling and polarimetry. The experimental data will then serve as further input to the simulations to properly prepare subsequent runs. Thus, depending on the results for SCT in the first run we then have three choices:
\begin{itemize}
	\item Further develop SCT at $T=140\,\mathrm{MeV}$ while using the same optical setting of COSY with super-periodicity P=2 and dispersion-free straight sections.
	\item Continue at the same or similar beam energy but change to an optical setting with higher super-periodicity P=6. This reduces the number of intrinsic resonances to one, but further boundary conditions for electron cooling and polarimetry with dispersion-free straight sections must be investigated.
	\item Use a different beam energy based on simulation results where intrinsic resonances are less harmful and $\Delta\nu_s\approx 0$. This will likely be a larger change in beam energy compared to the previous setting. Therefore, it is not possible to estimate beam time for this option at this point, as the implications for beam cooling and polarimetry would need to be evaluated based on the selected beam energy. 
\end{itemize}

While the development of long SCTs for deuterons was a continuous process over several years since 2012, the experience gained during that time should help to keep the necessary time for protons much shorter. \textbf{In total we estimate three separate runs with three weeks of beam time (plus two MD weeks) each} in case we can continue with the same or a similar electron cooler setup. A separate proposal has been submitted to CBAC \#13 asking for a first run within the next scheduling period\,\cite{jedi2021_e009}. 

\cleardoublepage

\subsection{Spin transparency experiments for proton polarization control at integer spin resonances}
\label{sec:spintransparency}

\begin{center}
	Y. Filatov$^1$, A. Kondratenko$^{1,2}$, M. Kondratenko$^{1,2}$ N. Nikolaev$^3$,
	A. Melnikov$^4$, Y.\,Senichev$^4$, A. Aksentyev$^4$, A. Butenko$^5$, E. Syresin$^5$, F.\,Rathmann$^6$, P\, Lenisa$^7$\\
	\vspace{0.2cm}
	(in cooperation with the JEDI collaboration)\\	
	\vspace{0.2cm}
	$^1$ \textit{Moscow Institute of Physics and Technology, Moscow, Russia}\\
	$^2$ \textit{Science \& Technique Laboratory “Zaryad”, Novosibirsk, Russia}\\
	$^3$ \textit{L.D. Landau Institute for Theoretical Physics, Chernogolovka, Russia} \\
	$^4$ \textit{Institute for Nuclear Research RAS, Moscow, Russia}\\
	$^5$ \textit{Joint Institute for Nuclear Research, Dubna, Russia}\\
	$^6$ \textit{Institut für Kernphysik, Forschungszentrum Jülich, 52425 Jülich, Germany} \\
	$^7$ \textit{University of Ferrara and INFN, 44100 Ferrara, Italy} \\
\end{center}

\begin{mdframed}
	\begin{enumerate}
		\item \textbf{Estimated number of MD weeks: \textcolor{red}{6}}
		\item \textbf{Estimated number of BT weeks: \textcolor{red}{9}}
	\end{enumerate}
\end{mdframed}
\vspace{0.3cm}

\subsubsection{Introduction}
A novel technique for the control of particle spins in storage rings was proposed\,\cite{proposal-DFG-ST}. It is called spin transparency (ST).  It constitutes an efficient, highly flexible method to control the beam polarization in storage rings and to manipulate the spin direction in real time during the experiments. It is a new way to bring experiments with polarized beams to a new level of precision and to overcome the spin experimental crisis\,\cite{Suleiman:2021whz}. 

When a racetrack accelerator is operated in the ST regime without snakes, the beam energy corresponds to an integer spin tune, so that any spin orientation is reproduced turn-by-turn. The magnetic lattice of the synchrotron becomes \textit{transparent} to the spin. This implies that the particles are in an integer spin resonance. The spin motion in such a situation is highly sensitive to small perturbations of the fields along the orbit. This high sensitivity threatens the stability of the beam polarization, but it allows implementing a simple and efficient spin control system. A spin navigator, which is a flexible device consisting of elements with weak constant or quasi-stationary fields rotating the spins about a desirable direction by a small angle, will be used. Strength of the applied spin navigator field must be larger than all small perturbative fields arising from magnetic structure imperfections and beam emittances to provide stable spin motion\,\cite{PhysRevAccelBeams.24.061001}.

The ST technique allows one to:
\begin{itemize}
	\item control the polarization by weak magnetic fields, not affecting the orbital dynamics,
	\item maintain stable polarization during an experiment,
	\item set any required polarization direction at any orbital location in a storage ring,
	\item change the polarization direction during an experiment,
	\item do frequent coherent spin flips of the beam to reduce experimental systematic errors, 
	\item carry out ultra-high precision experiments.
\end{itemize}

The basic ideas of the ST method were first formulated by the Russian team members in the process of the design of the figure-8 booster and collider synchrotrons of the Medium-energy Electron-Ion Collider (MEIC) project at Jefferson Lab\,\cite{Morozov:2015nfa}. Later on, the principle was applied in the design of the racetrack rings with two solenoidal snakes of the Nuclotron-based Ion Collider fAcility (NICA)\cite{Kovalenko_2017}. The ST mode of operation at the Relativistic Heavy Ion Collider (RHIC), the electron-Ion Collider (eIC) at Brookhaven, and the Electron-ion collider in China (EicC) has been recognized as being necessary\,\cite{Morozov:2015nfa,PhysRevAccelBeams.23.021001}.

The first-ever proof-of-principle experiment in ST mode should be conducted at COSY. It is a unique storage ring for spin physics with its excellent instrumentation and experience with polarized beams. In the foreseeable future there will be no competitors to COSY as a testing ground of ST and spin navigator ideas. 

Successful demonstration of ST approach opens up new possibilities for COSY in the field of unique experiments with polarized hadrons. The method can be applied to enhance the scientific potential of the semi-electric and all-electric storage rings for the search for electric dipole moments.

The project will be a natural new addition to the JEDI and CPEDM research programs. The obtained results will contribute to better understanding of proton spin dynamics and development of proton Spin-Coherence Time (SCT) investigations at COSY. This program will be an extension to the integer spin tune of the beam bump diagnostics of spin rotations by imperfection magnetic fields, currently under scrutiny by JEDI collaboration. The study is also relevant to Frequency domain method of EDM measurement in terms of experimental measurement of effective Lorentz factor $\gamma_\text{eff}$. That can also contribute to the design and construction of Prototype ring (PTR)\,\cite{cpEDM2021} and other accelerators for EDM measurements. 

The project is supported by DFG-RSF Joint German-Russian Scientific Cooperation (funding period 2022-2024)\cite{proposal-DFG-ST}. A detailed proposal for the experiment will be submitted for the CBAC session \#13\,\cite{proposal-Filatov}. For the investigations regarding the spin transparency mode, the following timetable is anticipated:

\subsubsection{Scientific Program 2021 -- 2024}

\paragraph{2021-2022:}
The principal task will be the estimation of technical parameters of the spin navigators for the manipulation of the in-plane polarization of protons in the COSY ring.

The collaboration studies in 2021-2022 will include:
\begin{itemize}
	\item theoretical evaluation of the strength of ST spin resonance at spin tunes equal to 2, 3 and 4 induced by ring imperfections and by orbital emittances;
	\item analysis of the impact of synchrotron oscillations on the spin dynamics;
	\item development of a description of spin-orbit coupling and spin-tracking simulations of the spin dynamics in the ST regime;
	\item formulation of technical specifications on magnetic elements of the COSY ring to be used as spin navigators
\end{itemize}

The experimental program in 2021-2022 will include:
\begin{itemize}
	\item measurement of resonance strengths during fast resonance crossings with different acceleration rates (no additional equipment is needed to be installed);
	\item adiabatic capture and delivery of polarized protons into the region of a spin resonance at $\gamma G = 2$ (existing \SI{2}{MeV} cooler solenoid will be used).  
\end{itemize}

\paragraph{2023:} 
The principal task will be the construction of a spin flip system for the ST regime: installation of a spin navigator in the ring and proof-of-principle test of the ST approach at COSY. 

The collaboration studies in 2023 will include:
\begin{itemize}
	\item analysis of the requirements on the spin navigator fields for the preservation of polarization during multiple spin flips;
	\item development of spin-orbit aspects of spin flip in the ST regime; 
	\item spin-tracking simulations of spin flips in the ST regime.
\end{itemize}

The experimental program in 2023 will include:
\begin{itemize}
	\item experiment on the spin-navigator driven stabilization of the in-plane polarization of protons at COSY with the analysis of the experimental data for optimization of the final ST and spin navigator experiment.
\end{itemize}

\paragraph{2024:} 
The principal task will be the experimental tests of the ST at integer spin tunes at COSY. 

The collaboration studies in 2024 will include:
\begin{itemize}
	\item numerical simulations of compensation of imperfection effects on the spin coherence time in the ST regime.
\end{itemize}

The experimental program in 2024 will include:
\begin{itemize}
	\item final experiment to test the ST regime with solenoidal spin navigators and  operation of the multiple spin-flip system in the ST regime at COSY;
	\item investigations of long-term stability and preservation of polarization;
	\item experiments to check the stability of the method by decreasing the fields in the solenoids to find the lowest acceptable fields in navigators.
\end{itemize}

\subsubsection{Beam time estimate}
2 weeks of MD and 3 weeks of BT will be requested for each year during the period 2022 -- 2024, i.e., in total 6 MD weeks and 9 weeks of beam time. Because of the complexity of setting up COSY for such a  non-standard operation at $G\gamma=2$, we believe the experiment needs to be performed independently from other studies, i.e., with a dedicated MD preparatory stage. More detailed estimates of the BT will be given in the main proposal\,\cite{proposal-Filatov}.

\cleardoublepage

\subsection{Spin-tune mapping with orbit bumps}
\label{sec:spin-tune-orbit-bumps}

\begin{center}
	A. Saleev$^1$, V. Shmakova$^1$\\
	\vspace{0.2cm}
		(for the JEDI collaboration)\\
	\vspace{0.3cm}
	$^1$ \textit{Institut für Kernphysik, Forschungszentrum Jülich, 52425 Jülich, Germany} \\
\end{center}

\begin{mdframed}
	\begin{enumerate}
		\item \textbf{Estimated number of MD weeks: \textcolor{red}{4} }
		\item \textbf{Estimated number of BT weeks: \textcolor{red}{8} }
	\end{enumerate}
\end{mdframed}

\subsubsection{Introduction}
The electric dipole moment (EDM) signal constitutes a rotation of the spin in the electric field. In an all magnetic ring (COSY), it is the motional electric field $\propto [\vec\beta\times\vec B] $ along the radial $x$-axis around which the EDM precesses. As such, an EDM contributes also to a constant tilt of the stable spin axis
\begin{equation}
\vec c = \vec e_y  + \xi_{\text{edm}}\vec e_x
\end{equation}

On the other hand, nonuniform in-plane magnetic fields tilt the invariant spin axis towards $x$ or $z$ (see Fig.\,\ref{fig:1}),
\begin{equation}
\vec c = c_y\vec e_y  + \left(\xi_{\text{edm}}+c_x^{\text{mdm}}\right) \vec e_x + c_z^{\text{mdm}} \vec e_z.
\label{eq:cxiedm}
\end{equation}
While $c_y \simeq 1$, the projections $c_{x,z}$ depend on the specific location along the beam path $s$, chosen to define the one-turn spin transfer matrix (see also discussion in ref.\cite{proposal-Artem}). In-plane magnetic fields have two origins: one is the radial focusing fields of the quadrupoles and vertical steerers to control the beam on a closed orbit. Another one is imperfection fields produced by the uncontrolled alignment errors of the magnets. The spin rotations in the in-plane fields are non-commuting with the spin rotations around the vertical field of the dipoles. This leads to complex dependence of $\vec{c}$ on $s$. However, unlike $c_{x,z}^{\text{mdm}}$, the EDM contribution to $\vec{c}$ is $invariant$ along the orbit. It gives possibility to disentangle the EDM and Magnetic Dipole Moment (MDM) effects if non-invariant part of $\vec{c}$ can be described.

\subsubsection{Selected results from JEDI Experiment {\tt E010}}
\begin{figure}[tb]
	\centering
	\includegraphics[width=0.8\textwidth]{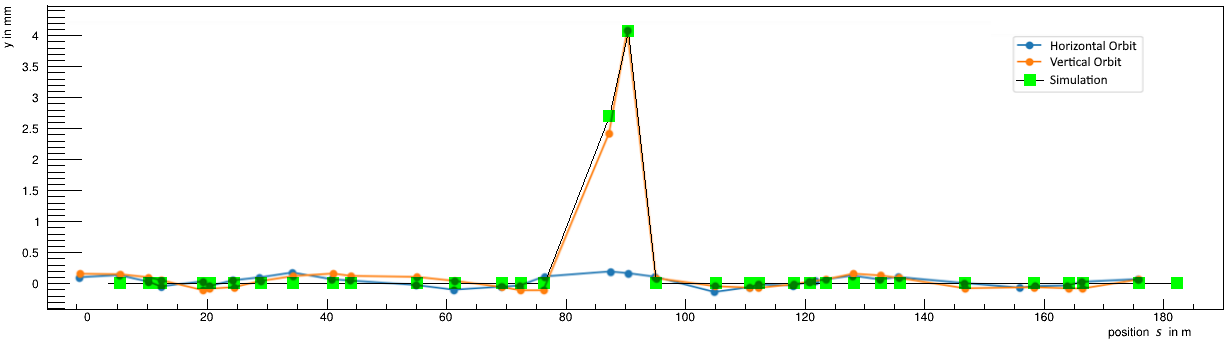}
	\caption{Modelled (green) ideal vertical orbit and measured (orange-vertical, blue-horizontal) orbit difference between 85 (no bump) and 115 (bump applied) seconds in the beam cycle. The proportionality coefficients for steerers to create closed bump were determined from the simulated and measured Orbit Response Matrices correspondingly.}
	\label{fig:2orbit} 
\end{figure}

In the JEDI experiment {\tt E010} in August -- September 2020, we tested a demerit of spin rotations, a "commutation failure", by creating a controlled closed orbit distortion over 1/8 part of the ring circumference - a vertical closed orbit bump. It was made by three steerers MSV18, MSV20 and MSV22 for a special period in time of the beam storage cycles (see measured orbit difference on Fig.\,\ref{fig:2orbit}). 
\begin{figure}[tb]
	\centering
	\includegraphics[width=0.8\textwidth]{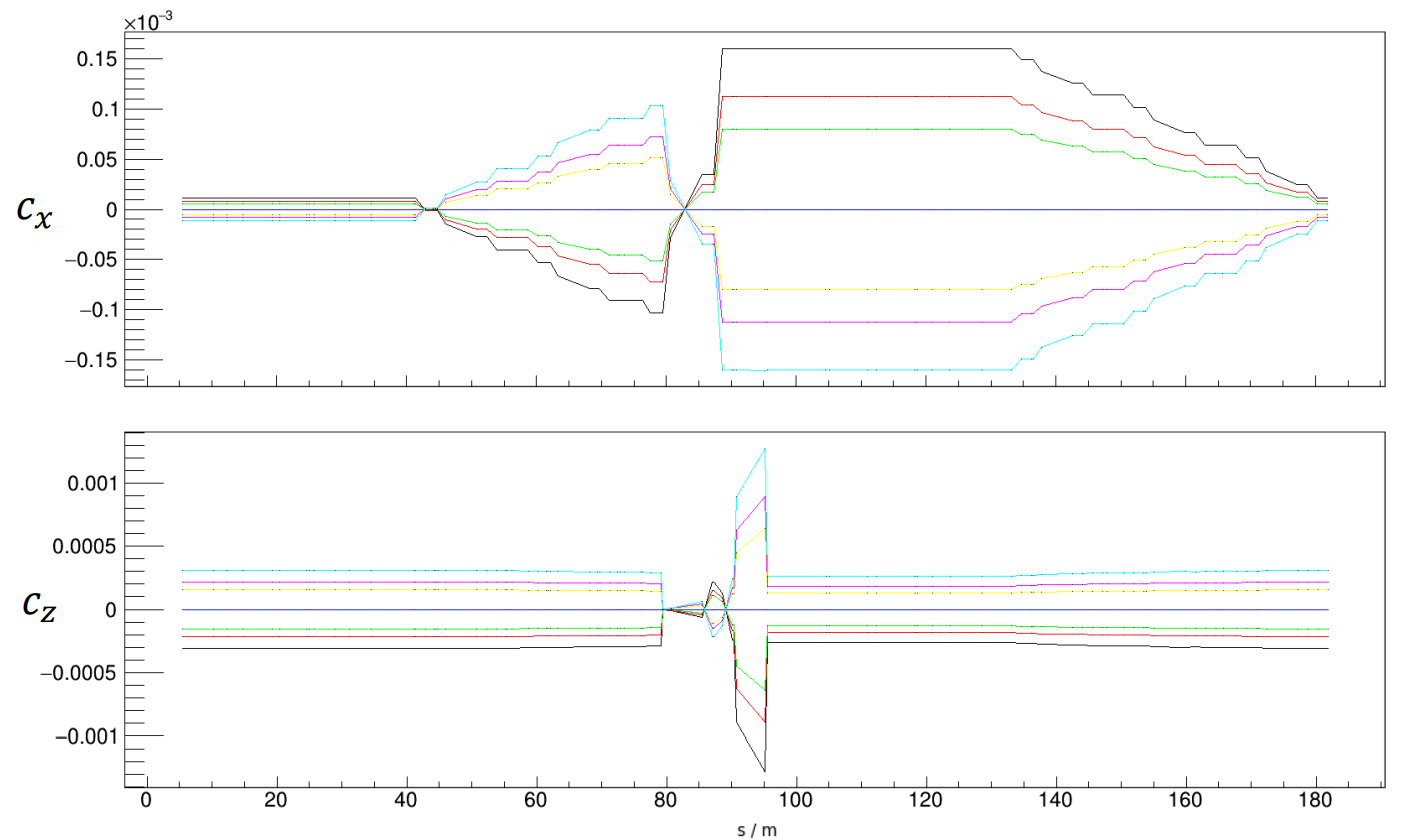}
	\caption{Dependence of the in-plane components of the invariant spin axis $\vec c$ along the closed orbit path $s$. (To track the closed orbit particle, beam and spin tracking package COSY-Infinity\,\cite{COSY-Infinity}was used.) Color code corresponds to the bump amplitudes that were used during the measurement, from $\SI{-4}{mm}$ (black), $\SI{-3}{mm}$ (pink) to $\SI{+3}{mm}$ (red), $\SI{+4}{mm}$ (cyan)}
	\label{fig:1} 
\end{figure}

An ideal model of COSY ring gives the prediction of $\vec{c}(s)$ (Fig.\,\ref{fig:1}) for all of the bump amplitudes (for example, orbit simulation shown on Fig.\,\ref{fig:2orbit}). In order to determine the dependence of projection $c_z$ on the bump amplitude in the experiment, we used a special method called ``spin tune mapping``\,\cite{PhysRevAccelBeams.20.072801}.  This method is based on the outstanding ability to determine the spin tune with a relative error of $\num{1e-10}$ during a \SI{100}{s} long beam cycle  at COSY from the time dependence of horizontal polarization\,\cite{PhysRevLett.115.094801}. For the same period of time in the cycles when the bump appears, two static solenoids, one in the target telescope, and one in the cooler telescope, were switched on. The spin tune was measured on the grid of solenoid currents $I_{1,2}$ applied at the fixed amplitude of the bump.

\begin{figure}[tb]
	\centering
	\includegraphics[width=0.8\textwidth]{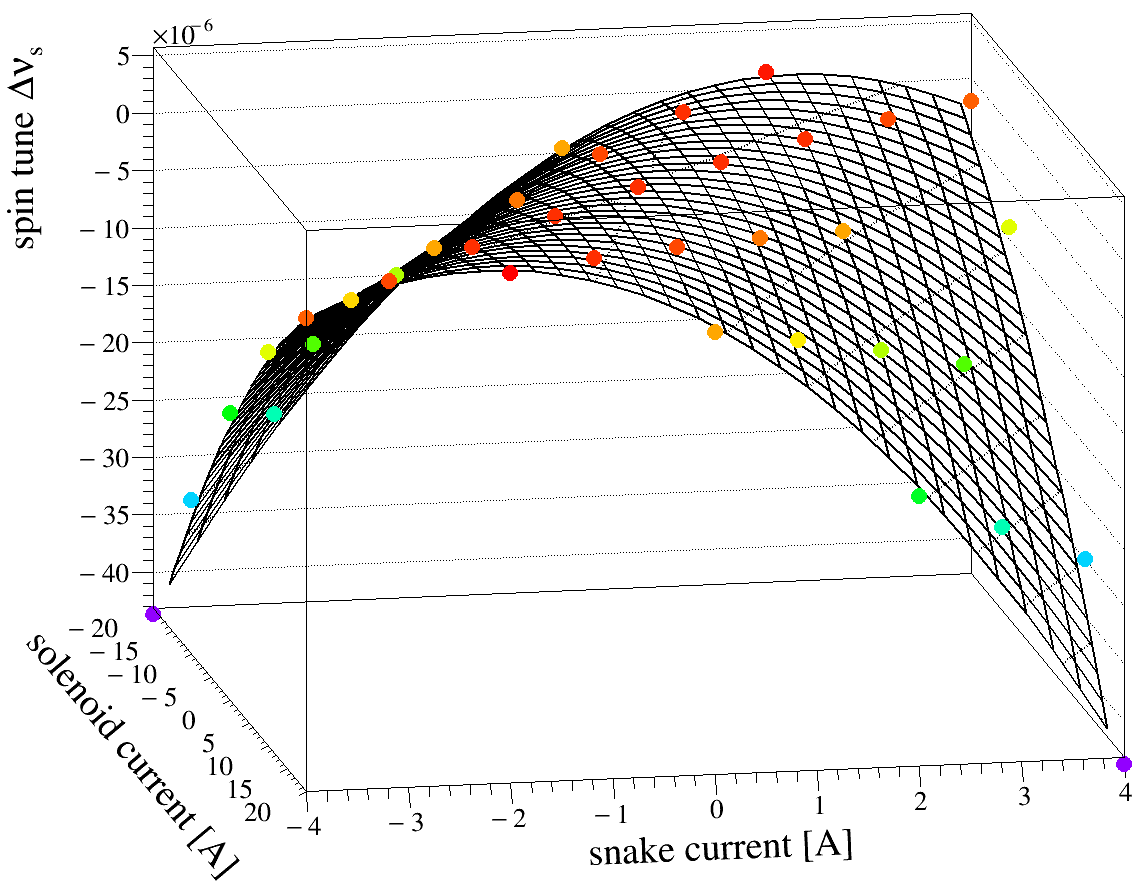}
	\caption{One of the spin tune maps measured at fixed bump amplitude $\SI{+3}{mm}$. Surface is a fit to data points (colored circles) by Eq.\,(\ref{eq:master}).}
	\label{fig:2} 
\end{figure}

Parabolic dependence of the spin tune shifts $\Delta\nu_s$, which are calculated as a change of spin tune relative to the baseline spin tune value $\nu_s$, given at the moment of time in cycle when the solenoid current and the bump amplitude were zero, fits non-lattice model where $c_z$  (at $s = \SI{16.27}{m}$ $c_z=c_\text{sol}$  for the 2\,MeV e-cooler compensation solenoid in target telescope and at $s = \SI{126.13}{m}$  $c_z=c_\text{snake}$ for superconducting snake solenoid at cooler telescope) and solenoid's current-to-spin-kick calibration $k_{1,2}$ are free parameters:
\begin{equation}
\begin{split}
-\pi\Delta\nu_s =  ( \cos a \cos b - 1  ) \cot\pi\nu_s - c_\text{sol} \sin a \cos b
- c_\text{snake} \cos a \sin b - \frac{\sin a \sin b}{\sin\pi\nu_s} \,,
\end{split}
\label{eq:master} 
\end{equation}
where
\begin{equation}
a = \frac{k_1\,I_1}{2} \quad \text{and} \quad b = \frac{k_2\,I_2}{2}\,.
\end{equation}
As a result of the spin tune mapping (see example on Fig.\,\ref{fig:2}), the values of $c_z$  are determined with angular precision $\sigma_{c_\text{sol}} = \SI{6.9}{\micro \radian}$ at the  2 MeV e-cooler solenoid and $\sigma_{c_\text{snake}} = \SI{3.6}{\micro \radian}$  at superconducting snake. The relative error on the spin tune shift $\Delta\nu_s$ is $\sigma_{\Delta\nu_s} = \num{3.7e-9}$. 

When  two static solenoids located at COSY telescopes are used, position dependence of $\vec{c}(s)$ is only partly uncovered. Nevertheless, fit results for spin tune maps at all of the measured bump amplitudes are in good agreement with the model prediction for dependence of $c_z$ projections at solenoids from the central steerer setting (see slope parameter $p1$ in Fig. $\ref{fig:solsnake}$). The values of central steerer  (MSV20) that correspond to the same amplitude of the bump in the model (in mrad) and measurement (in \%) were chosen as a reference ones. The settings for MSV18 and MSV22 were scaled accordingly to fulfill the condition of closed orbit bump, which is derived from the simulated (in case of model) and measured (in experiment) orbit response matrix.   It means as a matter of fact, that \emph{we created local orbit distortion by horizontal magnetic fields in the ring and described the resulting beam and spin dynamics}. Note that an offset parameter $p0$ at Fig.\,\ref{fig:solsnake} is non-vanishing in case of measured $c_\text{sol}$ and $c_\text{snake}$ due to the presence of alignment errors in the ring, which contribute to the tilt of invariant spin axis towards z-axis. 

The quantitative understanding of the local sources of the imperfection fields and their active compensation is indispensable for disentangling the EDM effect from the EDM-like background from interactions of the vastly larger magnetic dipole moment with the in-plane magnetic fields. 

\subsubsection{New proposal}
The new proposal aims at probing the local properties of the focusing fields at COSY. Spin tune maps with all available  vertical three-steerer bumps in the arcs should be measured and compared to the model (12 bump configurations in total). Developed method is applicable in the future storage rings as a tool for diagnostics of beam and spin dynamics when approaching frozen spin condition. At prototype EDM ring, it can be applied at \SI{30}{MeV} counter-circulating protons, to verify the achievements of the beam and spin-dynamic studies at pure magnetic rings with non-frozen spin. It is an important connecting step to test the model predictions for pure electrostatic lattice, preceding the measurements at strictly frozen spin condition.

The beam development requires setting up polarized electron-(pre)cooled deuteron beam at the momentum \SI{970}{ MeV/c}, measurement of orbit response matrix,delivering beam on the polarimeter target and testing the performance of an active shunts that will be connected to all of MXS family sextupoles to being able to suppress the current in a single sextupole. After the routine spin coherence time optimization with sextupoles that takes \SI{2}{\day} (two optimizations are expected during the beam time), a complete measurement of full spin tune map for one bump configuration takes \SI{2}{\day}, and \SI{24}{\day} in total for all bumps. In this regard, we would like to request \SI{2}{wk} of machine development and \SI{4}{wk} for measurements.

\begin{figure}[tb]
	\centering
	\includegraphics[width=0.8\textwidth]{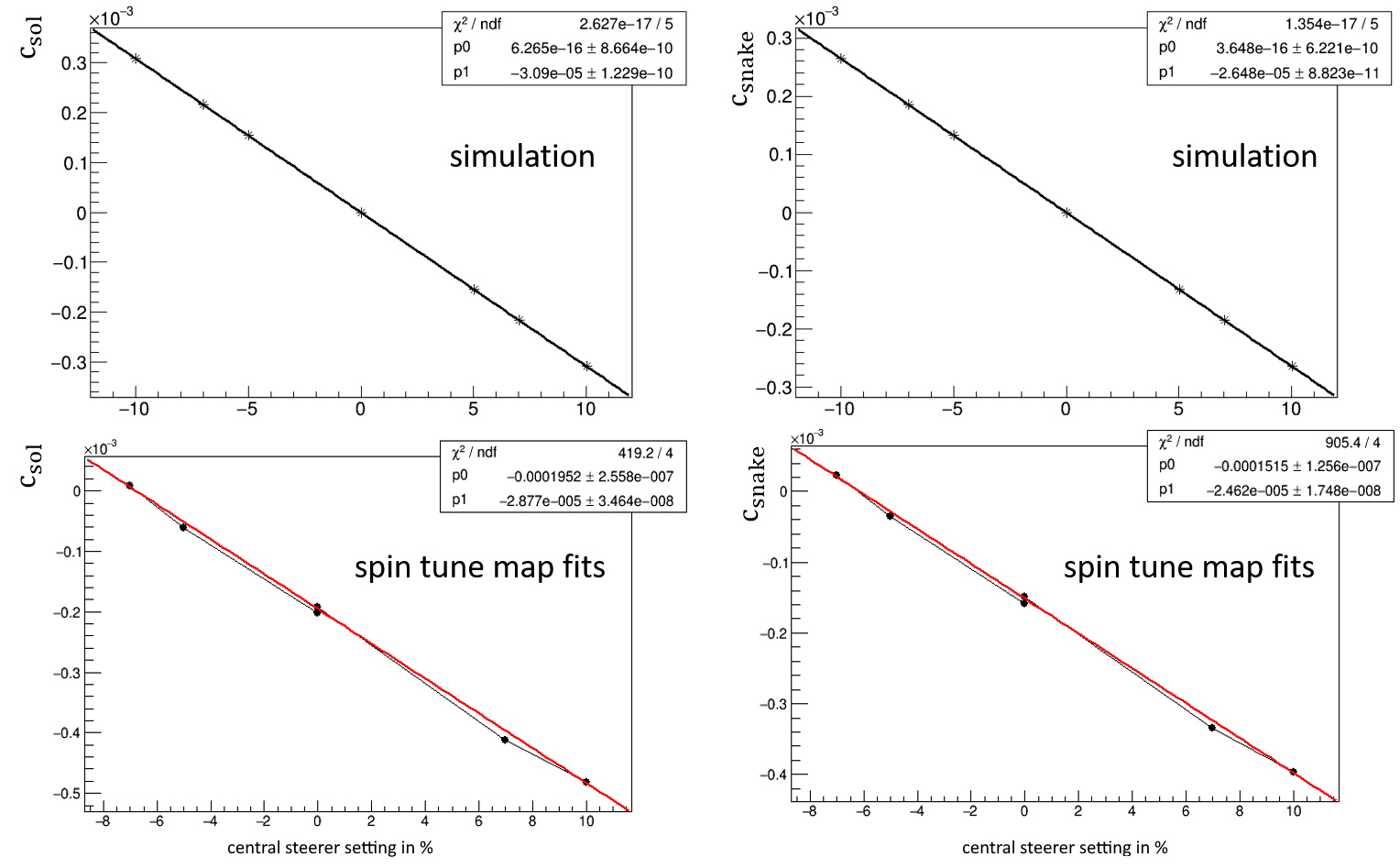}
	\caption{Dependence of the fit parameters $c_\text{sol}$ and $c_\text{snake}$ on the steerer setting and comparison to simulation results. Value $+10\%$ corresponds to the bump amplitude of $\SI{-4}{mm}$.}
	\label{fig:solsnake} 
\end{figure}

After the successful demonstration of the ability to take control over spin and orbital dynamics for all vertical 3-steerer bumps at COSY, a spin-tune mapping scheme with a global vertical orbit correction will be scrutinized. This would allow us to witness the impact of the applied vertical orbit correction at COSY on the observed tilt of the invariant spin axis (the measured parameter $p0$ in Fig.\,\ref{fig:solsnake} for the $z$-projection of $\vec{c}$). The required amount of beam time for this investigation is estimated to be 2 wk for machine development and 4 wk of beam time. Finally, the studies should be repeated for protons, once a sufficiently long spin-coherence time is achieved in the JEDI experiment on "Spin-coherence time investigations for protons", described in Sec.\,\ref{sec:spico-for-protons}. 

\cleardoublepage


\subsection{Axion search}
\label{sec:axion}

\begin{center}
	S.\,Karanth$^1$, J.\,Pretz$^{2,3}$, E.\,Stephenson$^4$\\
	(for the JEDI collaboration)\\
	\vspace{0.2cm}
	$^1$ \textit{Marian Smoluchowski Institute of Physics, Jagiellonian University, 30-348 Kraków, Poland}\\
	$^2$ \textit{Institut für Kernphysik, Forschungszentrum Jülich, 52428 Jülich, Germany} \\	
	$^3$ \textit{III. Physikalisches Institut B, RWTH Aachen University, 52056 Aachen, Germany} \\
	$^4$ \textit{Indiana University Center for Spacetime Symmetries, Bloomington, IN 47405, USA} \\
\end{center}

\begin{mdframed}
	\begin{enumerate}
		\item \textbf{Estimated number of MD weeks: \textcolor{red}{8} }
		\item \textbf{Estimated number of BT weeks: \textcolor{red}{52}}
		\footnote{\label{footnote:2.4}The 52 weeks of running corresponds to \SI{e7}{\second}. This is meant as a possibility that if another experiment claims the discovery of an axion/ALP in the mass range accessible at COSY one could verify/falsify the claim repeating measurements at a fixed frequency (for more details see text).}
	\end{enumerate}
\end{mdframed}

The Standard Model (SM) of particle physics provides no solution to the Dark Matter (DM) problem. As a consequence new particles outside the realm of the SM have been proposed as dark matter candidates. Among them are so called WIMP (Weakly Interacting Massive Particles). In spite of enormous experimental effort, no DM particle was found up to now. It it thus important to widen the search to other, theoretically well motivated DM candidates\,\cite{DeMille:2017azs}. These include ultralight particles like axions and axion-like particles (ALPs). Originally axions were introduced to solve the so called strong $CP$-problem in the Standard Model of particle physics\,\cite{PhysRevLett.38.1440,PhysRevD.16.1791,PhysRevLett.40.223,PhysRevLett.40.279}. If axions are responsible for dark matter, they must be very light with their mass being below \SI{e-6}{eV}, i.e., 15 orders of magnitudes lighter than the proton.

Axions may interact with ordinary matter in different ways. Most of the axion searches are based on the axion-photon interaction. Other experiments make use of the axion interaction with atomic nuclei or the coupling to the spin. To study the latter effects, the well known nuclear magnetic resonance technique\,\cite{PhysRevX.4.021030} can be used. 

Storage rings offer the possibility to look at the effect of axions on the spin motion of a particle ensemble\,\cite{Semertzidis:2019gkj,Pretz2020}. Axions couple to gluons inducing an oscillating electric dipole moment (oEDM) on nucleons or nuclei\,\cite{PhysRevD.84.055013}. These oEDM can be detected in a storage ring by making use of resonant methods enhancing the effect and allowing for unprecedented precision.

\begin{figure}[htb]
	\centering
	\includegraphics[width=0.6\textwidth]{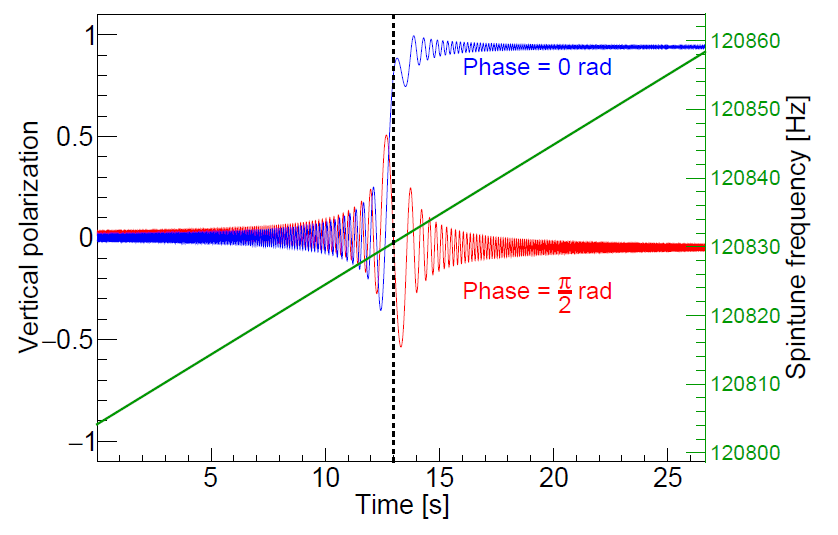}
	\caption{Example calculation where the spin tune frequency (green) was scanned in search of the resonance. The signal is the jump in the vertical polarization at the resonance crossing (black dashed line). The blue and red curves represent phases of 0 and  $\pi/\SI{2}{\radian}$ between the oscillating EDM and the rotating in-plane polarization. \label{fig:jump}}
\end{figure}

In the spring of 2019, the JEDI Collaboration conducted a successful proof-of-principle experiment at COSY to search for axions using a \SI{0.97}{ GeV/c} deuteron beam polarized in the ring plane. The oscillating EDM, which is aligned along the spin direction, experiences a torque due to the particle frame electric field from the bending magnets that rotates the beam polarization in a vertical plane. If the axion oscillation frequency matches the spin tune frequency, then there will be a resonant buildup of vertical polarization.
As the axion frequency is not known, the spin tune frequency is ramped in the search for a signal that is a jump in the vertical polarization, as seen in the Fig.\,\ref{fig:jump} model calculation.
\begin{figure}[!]
	\centering
	\includegraphics[width=0.6\textwidth]{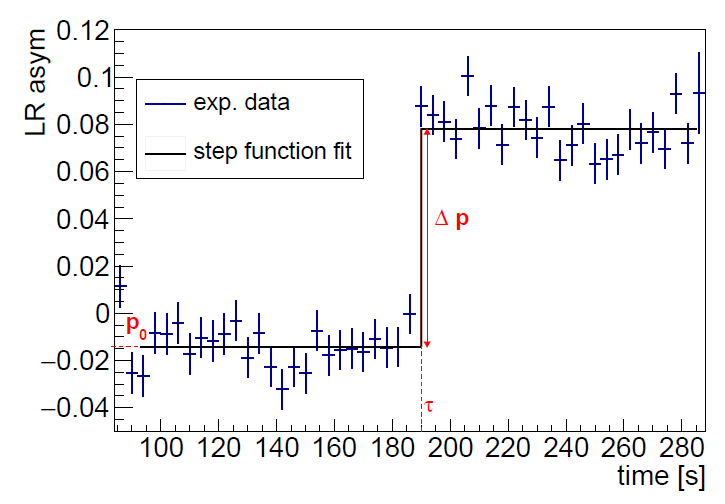}
	\caption{Polarization jump seen at resonance crossing in the case of an RF Wien filter scan. Also shown is the step function used to model the size, $\Delta p$, and time, $\tau$, of the jump. \label{fig:jump2}
	}
\end{figure}

To avoid accidental cancellation of the signal because of a problem with the axion phase (red curve), four beam bunches were run simultaneously with adjacent bunches having perpendicular in-plane polarization directions. To verify that the setup was capable of detecting and axion-like resonance, a test signal was made using an RF Wien filter with a radial magnetic field\,\cite{Slim:2016pim}. When on resonance with the rotating polarization at a frequency of $(1-\gamma G) f_\text{rev}$, a driven oscillation in the vertical polarization is produced that appears as jumps for each of the four beam bunches, as shown in Fig.\,\ref{fig:jump2} for one bunch. The size of the jump is confirmed by model calculations. This confirms the calibration of the sensitivity
to the size of the axion-nucleon coupling.

\begin{figure}[htb]
	\centering
	\includegraphics[width=0.7\textwidth]{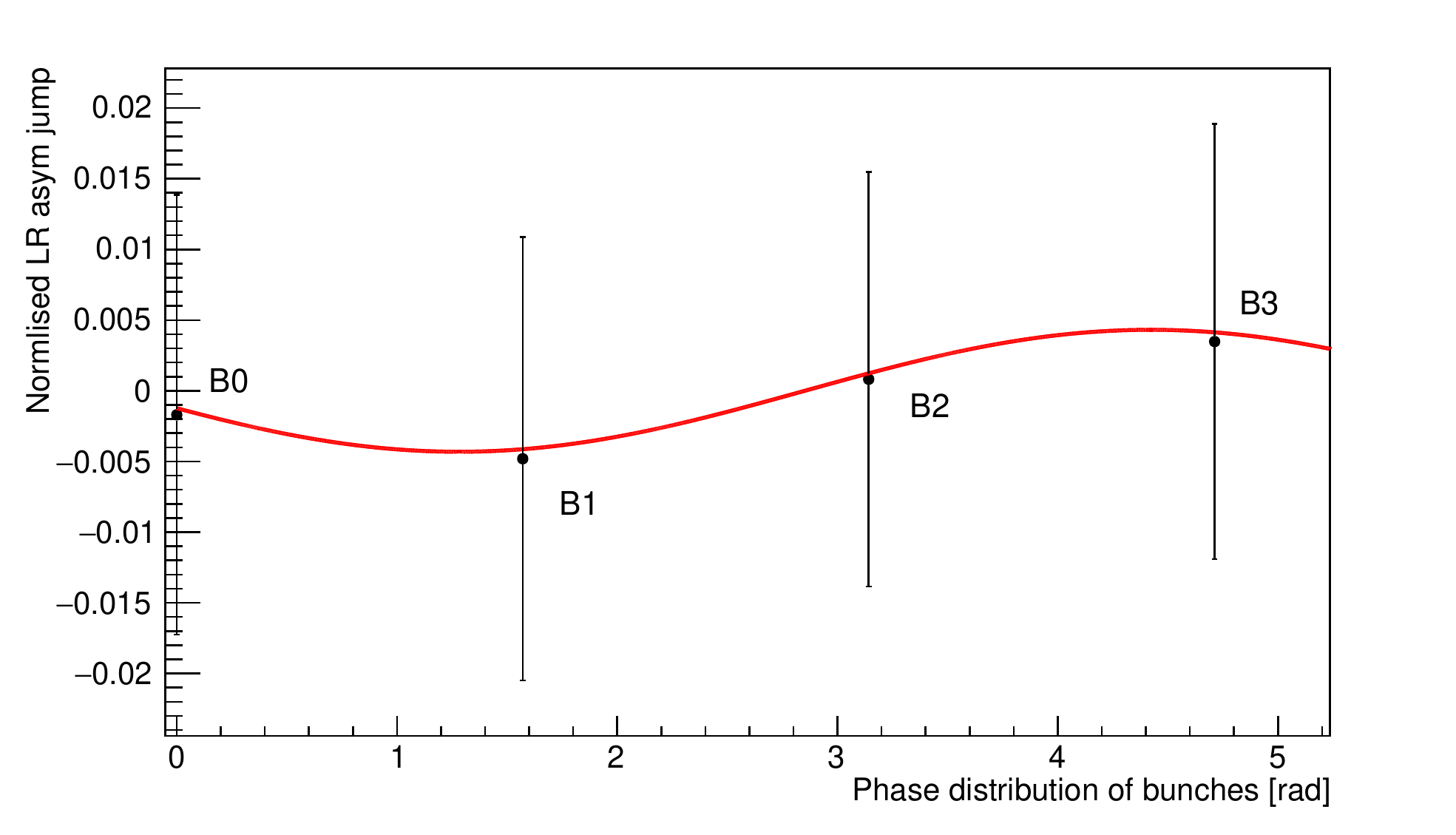}
	\caption{Jump sizes for the four beam bunches assuming the jump occurred in a particular time bin. The errors represent the contributions from the beam polarimeter as shown in figure~\ref{fig:jump2}. The red curve is a sinusoidal best fit to these data.
		\label{fig:4ampl}}
\end{figure}

\begin{figure}[!]
	\centering
	\includegraphics[width=0.7\textwidth]{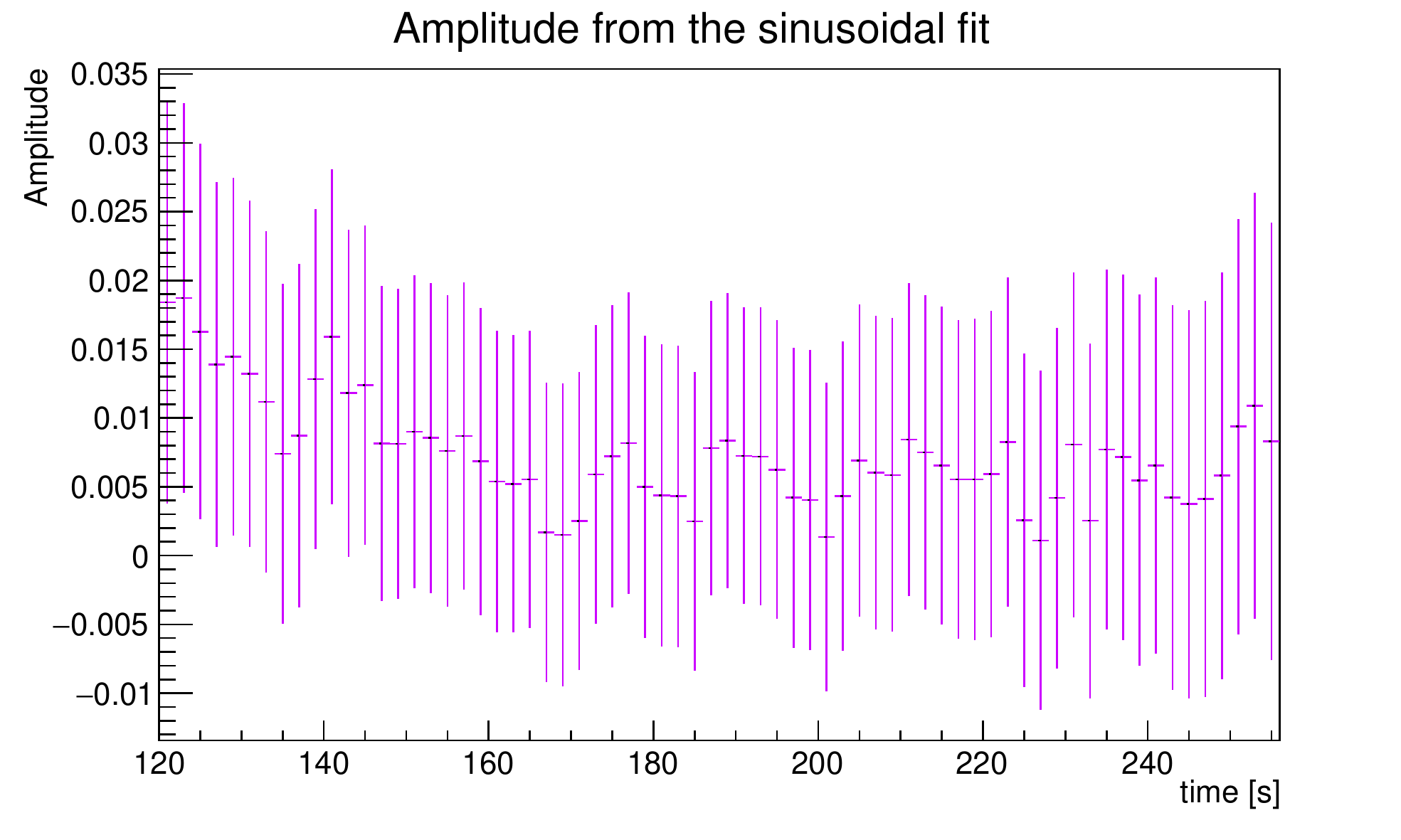}
	\caption{Values of the amplitude of the best fit sine wave for each time bin.
		All amplitudes are shown as positive. All of the asymmetry data contributes to each error bar shown. Nearby amplitudes are correlated, as appears in the amplitude slope forward of \SI{160}{s}.	The scanned frequency range is \SI{120833.0}{\hertz} to \SI{120849.6}{\hertz}. \label{fig:amplitude}}
\end{figure}

Figure\,\ref{fig:jump2} also illustrates how the step function may be used to survey data from the axion scans in order to locate a possible resonance. For each time bin and beam bunch, a value of $\Delta p$ is calculated. An example for a single time bin is shown in Fig.\,\ref{fig:4ampl}. The amplitudes of the the red curve enters Fig.\,\ref{fig:amplitude}.%

Figure\,\ref{fig:amplitude} contains the amplitudes and errors for each time bin in a single frequency scan. Each point is calculated from the same set of asymmetry measurements for a scan with four beam bunches. An increase in error size during the scan reflects a loss of in-plane polarization (all data are normalized to a unit polarization as the scan starts). In such a data set where the amplitudes reflect a sum in quadrature over neighboring bunch asymmetries, the amplitudes will always be positive, even if the expectation for a signal is zero. This bias complicates the assignment of a meaningful upper limit; Bayesian approaches are currently being developed. When completed, a limit will be available for about \num{e6} adjoining frequency scans covering a range from \SI{119.997}{\kilo \hertz} to \SI{121.457}{\kilo \hertz}, or an axion mass range of  $(4.95-5.02) \times \SI{e-9}{\electronvolt}$.

\begin{figure}[htb]
	\centering
	\includegraphics[width=0.5\textwidth]{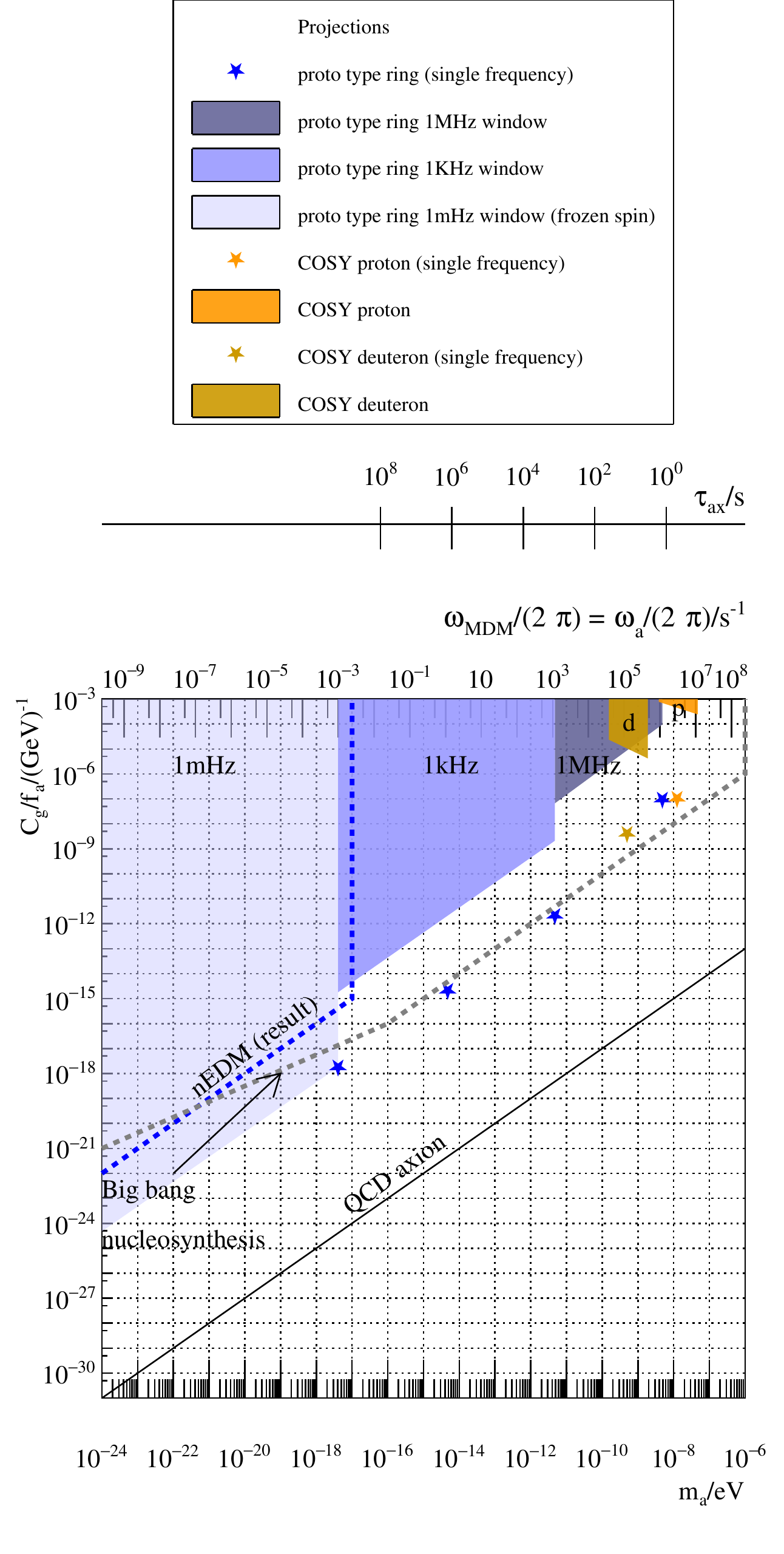}
	\caption{\label{fig:axion_range}
		One $\sigma$ limits for the axion-gluon coupling $C_g/f_a$ reachable within one year running at a fixed frequency (stars) or over a given frequency range (areas) for COSY (orange) or the prototype ring (blue). A year of running corresponds to \SI{e7}{\second}  seconds. A beam intensity of \num{e9} particles per cycle was assumed
		for COSY running. (Figure adapted from\,\cite{Pretz2020}.)
	}
\end{figure}

In future applications, this technique may be used to survey a much broader range of frequencies (masses) or, by repeating scans in a limited range, improve the sensitivity of the search by as much as $10^5$, as suggested in Fig.\,\ref{fig:axion_range}.

To be useful, beam times on the order of a year (about \SI{e7}{\second} considering setup and overhead) might be required. This would represent a substantial commitment of personnel time and laboratory resources. It should only be considered in the event that outside physics arguments or observations (e.g., a claimed axion discovery) create a well-defined opportunity. Otherwise, the authors will bring the analysis to closure and publish the search method in the literature as the final results become available. In the event that axion signals in such a storage ring experiment are easily discernible, long-term observations could generate data relevant to the observational frequency width of axions bound to the galaxy or density fluctuations that appear as changes in the coupling strength to the axion field, quantities that may have cosmological implications.

\cleardoublepage

\subsection{Interplay of spin-coherence time and stochastic cooling}
\label{sec:cooling}

\begin{center}
	B.\,Breitkreutz$^1$, N.\,Shurkhno$^1$, and R.\,Stassen$^1$ \\
	(for the IKP-4 group)\\
	\vspace{0.3cm}
	$^1$ \textit{Institut für Kernphysik, Forschungszentrum Jülich, 52425 Jülich, Germany} \\
\end{center}

\begin{mdframed}
	\begin{enumerate}
		\item \textbf{Estimated number of MD weeks: \textcolor{red}{--} \footnote{\label{note1} Project not yet in a state that would allow beam time estimates.}}
		\item \textbf{Estimated number of BT weeks: \textcolor{red}{--} $^a$}
	\end{enumerate}
\end{mdframed}

\subsubsection{Introduction}

The EDM precursor experiment in COSY uses polarized deuteron and in the next step proton beams at low momenta below 1\,GeV/c, that are stored for several minutes. In order to increase the spin coherence time, beam cooling is necessary. Electron cooling is applied to pre-cool the beam, but the solenoids and toroids of the electron cooler are not perfectly compensated.

Due to the magnetic fields in the electron beam line of the electron cooler leading to large orbit deviations in the range of tens of mm of stored beams, systematic effects in the EDM ring would be significantly increased. In addition, two electron coolers would be needed for CW and CCW beams in the final EDM  storage ring, which would also be unacceptable in terms of the cooling-section length ($< \SI{10}{m}$ cooling section) to achieve the required cooling forces. Therefore, stochastic cooling is the only option for an EDM storage ring. Unfortunately, the existing stochastic cooling system at COSY is not sensitive at low beam velocities ($\beta < 0.75$). New pickup and kicker structures need to be built and tested in COSY to investigate the cooling forces at low beta ($\beta \approx 0.3$ for PTEDM and $\beta \approx 0.6$). Additionally, systematic effects on the spin motion induced by the broad-band kicker system needs to be studied to get an estimate for systematic spin rotations.

\subsubsection{System design}
Additional work on stochastic cooling pickups and kickers for a system dedicated to low beam velocities of approximately \SI{0.5}{c} was performed. The design is based on the slot-ring type pickups that have been developed for the High Energy Storage Ring (HESR)\,\cite{SC1:Stassen-2017}, but optimized for slow particles and a low frequency band of $350$ to $\SI{700}{MHz}$. Since the structures get much bigger in comparison to the HESR version, mechanical properties must be reconsidered and a trade-off between electrical properties, cooling performance and manufacturability must be found.
\begin{figure}[ht]
	\begin{center}
		\includegraphics[width=0.5\textwidth]{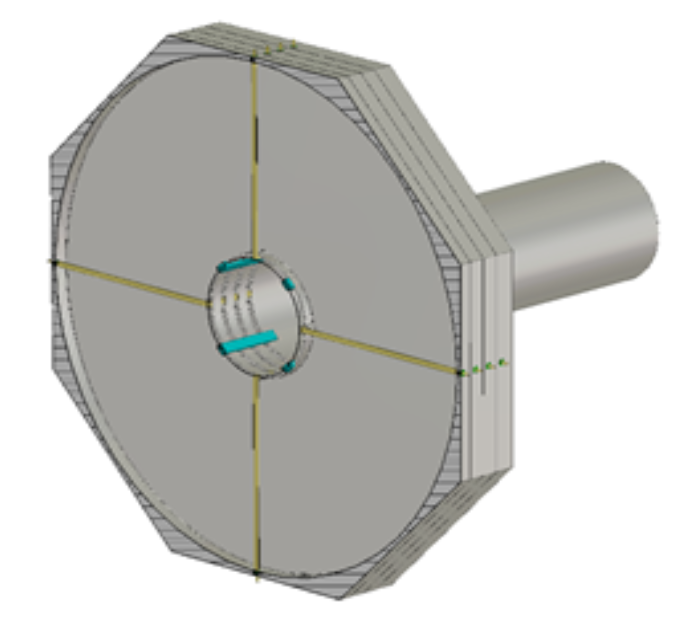}
		\caption{Proposed pickup and kicker hardware for low beta beams.}
	\end{center}
	\label{SC_PUPKICKER}
\end{figure}

Starting from the HESR geometry, the slot-rings were optimized for a high longitudinal shunt impedance. Therefore, kickers have been simulated with CST Microwave Studio. By varying different parameters like the slot-width and the cell height, a geometry with a sufficiently high impedance in the desired frequency band was found\,\cite{SC2:Breitkreutz-2017}. Figure\,\ref{SC_PUPKICKER} shows a cut through the transverse plane of the final structure. 

\begin{figure}[ht]
	\begin{center}
		\includegraphics[width=0.5\textwidth]{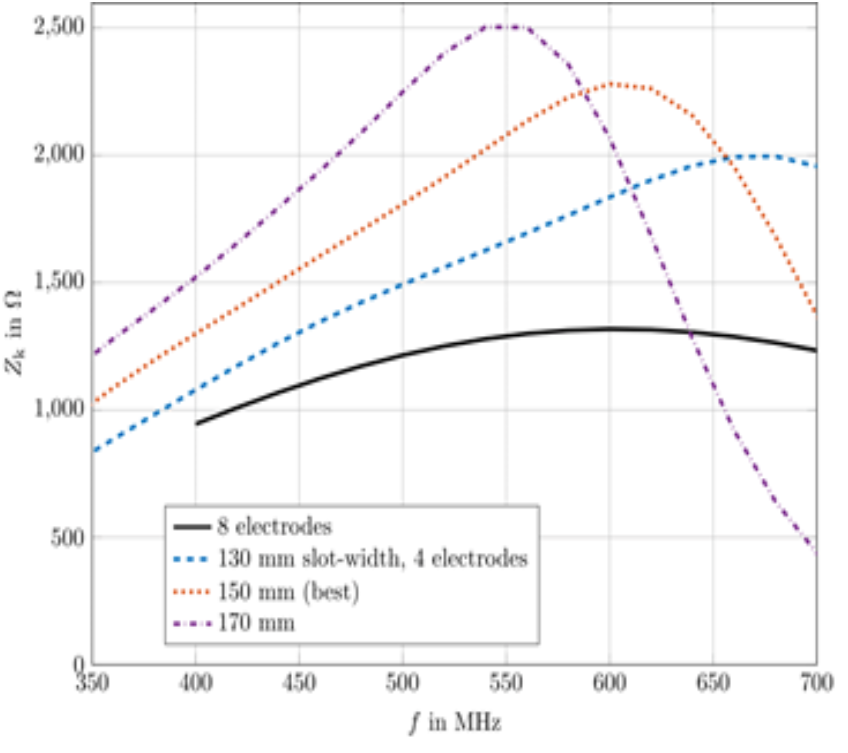}
		\caption{Longitudinal shunt impedance of different 64-cell kickers.}
	\end{center}
	\label{SC_IMPEDANCE}
\end{figure}

In Fig.\,\ref{SC_IMPEDANCE}, the shunt impedance of the final design is shown and compared to some variations. The most sensitive parameter is the slot-width, which tunes the optimum frequency in the desired band. Since this leads to very huge structures, some mechanical properties had to be adjusted. The thickness of the base plates was in-creased for a better stiffness. Supports made of PEEK have been inserted between the single rings to fix the distance, and sleeves in the coaxial connectors center the long electrodes.
Furthermore, the number of electrodes per ring was reduced from eight to four. More electrodes increase the bandwidth of the structure, but reduce the peak sensitivity. Nevertheless, as can be seen in Fig.\,\ref{SC_IMPEDANCE}, four electrodes still lead to higher impedances along the whole band.

Although the structure was optimized for longitudinal performance only, the transverse shunt impedance turned out to be better than \SI{1600}{\ohm} in the whole band. This is sufficiently large for an efficient transverse cooling.

\begin{figure}[ht]
	\begin{center}
		\includegraphics[width=0.5\textwidth]{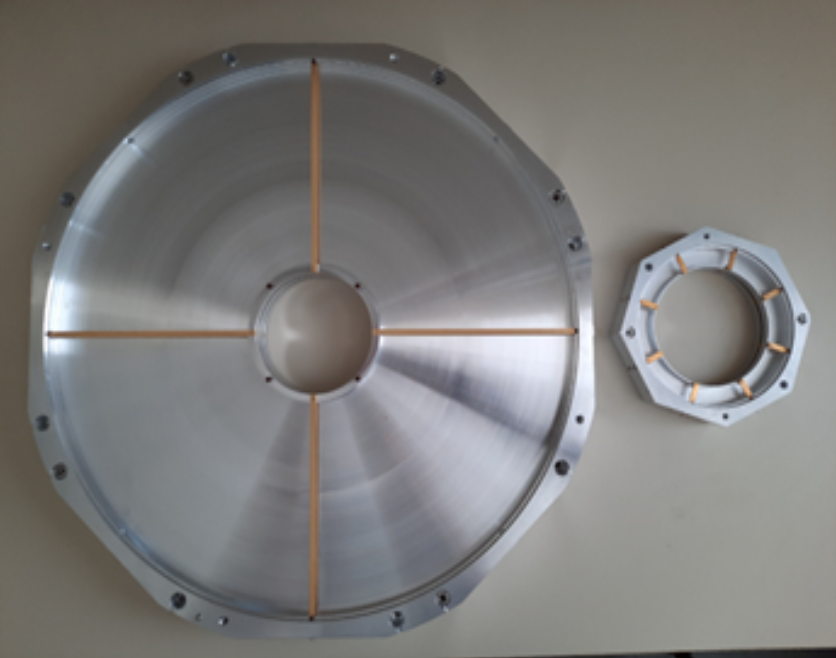}
		\caption{First EDM-ring (left) compared with HESR-ring (right).}
	\end{center}
	\label{SC_RING}
\end{figure}

First prototype rings were fabricated to verify the machining process (see Fig.\,\ref{SC_RING}). The structure is comparatively huge with a diameter of approximately \SI{40}{cm}. The stiffness was adjusted by increasing the thickness of the base plates to \SI{2}{mm}, and adding plastic supports between the rings and around the electrode outlets. Simulations showed a negligible influence on the electromagnetic properties.
\begin{figure}[ht]
	\begin{center}
		\includegraphics[width=0.4\textwidth]{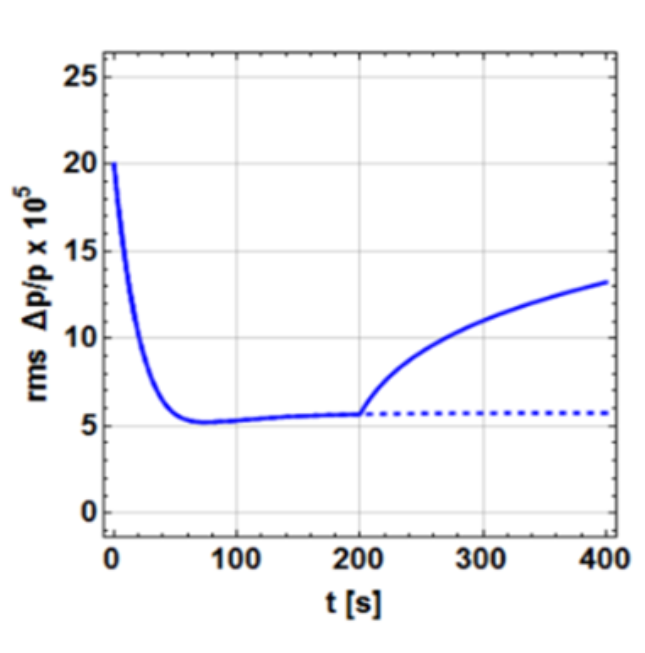}
		\includegraphics[width=0.4\textwidth]{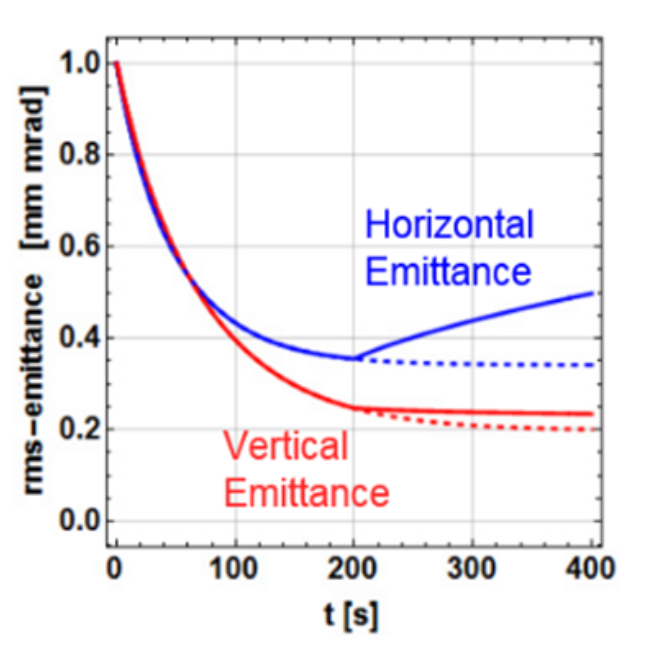}
		\caption{ Simulation results of momentum spread (left) and emittances (right) during cooling. After \SI{200}{s}, cooling is turned off.}
	\end{center}
	\label{SC_SIMULATION}
\end{figure}

\subsubsection{Beam simulation}

The cooling performance of the proposed system was simulated for an EDM beam. For the simulations, intra-beam scattering was considered. For the longitudinal case, filter cooling was assumed. Figure\,\ref{SC_SIMULATION} shows how the rms momentum spread and the rms emittances evolve in 200 seconds of cooling. The longitudinal time constant is approximately 30\,s, the transverse \SI{80}{s}. The equilibrium states are \num{6e-5} for momentum spread, and $0.35/\SI{0.2}{mm.mrad}$ for horizontal and vertical emittance.

After \SI{200}{s}, the cooling is turned off. The beam grows again due to intra-beam scattering. Thus, instead of pre-cooling only, an operation during the whole experiment would be desirable. We already investigated the influence of stochastic cooling on polarized beams in the past at higher energies, and could not observe any depolarizing effects\,\cite{SC3:Stockhorst-2013}. Therefore, we are hopeful that cooling during the experiments is possible.

Another important result of the simulation is the amount needed RF power. It turned out that off-the-shelf \SI{5}{\watt} power amplifiers are sufficient for the system. This is very beneficial, since power amplifiers are usually an important matter of expense in stochastic cooling systems.

The concept of signal routing has to be revised, as the low frequencies lead to very large combiner boards. First CST simulations using a ceramic tube between beam and structures look very promising. In this case the structures and combiners are outside the vacuum and a variety of different combiner concepts can be used.

Another important structure for stochastic cooling of low beta particles is a Faltin traveling wave structure\,\cite{SC4:Faltin-1978}. Further simulations will be carried out over the next few years to find out which solution is best suited for a dedicated EDM ring and which restrictions exist for both structures with regard to the beta value. Faltin structures do not require combiner boards and can be used in both directions, which can be interesting for cooling bunched clockwise and counter-clockwise beams in the same beam pipe.

\subsubsection{Conclusion and Outlook}

Stochastic cooling at low beta ($\beta < 0.5$) is scientifically extremely interesting and very important for various storage ring projects, e.g., EDM measurements of charged particles, spin filtering studies to polarize antiproton beams, and other possible precision experiments at low energies at GSI/FAIR. The effect of cooling on the beam motion (6D emittance) and the impact on the spin motion at low energies should be studied in detail at COSY. At this stage, there are still open questions regarding budget and personnel for this project and it remains to be seen if such a system will be operational at COSY by 2024. Therefore, no estimation for beam time requirements at COSY can be made at this point. The project could be significantly advanced by shifting priorities within TransFAIR.
\cleardoublepage

\subsection{Longitudinal spin filtering with protons}
\label{sec:spinfiltering}

\begin{center}
	P. Lenisa$^1$, A. Pesce$^2$, A. Nass$^2$, N. Nikolaev$^3$, and F. Rathmann$^2$\\
	\vspace{0.2cm}
	(for the PAX collaboration)\\
	\vspace{0.2cm}
	$^1$ \textit{University of Ferrara and INFN, 44100 Ferrara, Italy} \\
	$^2$ \textit{Institut für Kernphysik, Forschungszentrum Jülich, 52425 Jülich, Germany} \\
	$^3$ \textit{L.D. Landau Institute for Theoretical Physics, Chernogolovka, Russia} \\
\end{center}

\begin{mdframed}
	\begin{enumerate}
		\item \textbf{Estimated number of MD weeks: \textcolor{red}{7} }
		\item \textbf{Estimated number of BT weeks: \textcolor{red}{11} }
	\end{enumerate}
\end{mdframed}

\subsubsection{Objectives}
Following an unsuccessful attempt to polarize protons by spin flip \,\cite{OELLERS2009269}, the PAX collaboration has successfully performed a spin-filtering measurement with protons at the COSY ring by using a transverse polarized hydrogen target\,\cite{PhysRevSTAB.18.020101, AUGUSTYNIAK201264}. These studies were carried out in the framework of an ERC grant\,\cite{pax:ERC}. The COSY measurement is actually a determination of the transverse spin-dependent polarization build-up cross section and proves that spin-filtering can be considered as a method to polarize a stored beam \cite{Barone:2005pu}. PAX can use the unique environment offered by the COSY ring to transfer this method to longitudinal polarization. This point is motivated by the fact that all the different models for spin-filtering predict a significantly higher degree of polarization for the longitudinal case than for the transverse one. For the optimization of the polarization buildup it is therefore necessary to study both cases in conjunction. In addition, this approach represents the only way to obtain the relevant spin-dependent cross-sections for producing polarized antiprotons. Afterwards the PAX collaboration is ready to perform the corresponding experiments with antiprotons at FAIR\,\cite{Kacharava:2005wz}.

\subsubsection{Implementation}
The natural direction of the polarization in a storage ring is vertical with respect to the beam momentum; longitudinal polarization requires the introduction of a dedicated magnet system: namely a Siberian snake by which the spin-closed orbit at the target installed in the opposite straight section is oriented along the longitudinal direction \cite{Lehrach:2001db}. A Siberian snake has been installed in COSY, but still needs commissioning at the energy of the planned spin-filtering test (i.e., at a beam kinetic energy of $T \approx \SI{120}{MeV}$). The spin-filtering test will require also the installation of a couple of skew-quadrupoles (already existing) in COSY to compensate for the coupling effect of the snake and to allow for the beam-lifetimes required by a spin-filtering test. In preparation of the spin filtering test, the PAX detector \cite{Lenisa:2019cgb} has also to be installed at the PAX  interaction point and fully commissioned. The detector system is compatible both with the longitudinal spin-filtering experiments at COSY and the following experiments with antiprotons. The rest of the experimental apparatus necessary for the experiment: i.e., the polarized atomic beam source\,\cite{Nass2003633} and the target polarimeter \cite{BAUMGARTEN2002606, BAUMGARTEN2003268}, the target chamber and the vacuum system has been already developed and commissioned. 

\subsubsection{Measurements / Beam requests}
\begin{enumerate}
	\item {\bf Commissioning run for the Siberian snake with the skew-quadrupoles}
	\begin{itemize}
		\item After installation {\bf 2 weeks + 2 MD} are needed, operational earliest in 2022/23.
		\item  Mile	stones: long beam life time, long polarization life time.
	\end{itemize}
	\item {\bf Commissioning run for the polarized target and detector system}
	\begin{itemize}
		\item Following the success of step 1., and after the preparation of the polarized
		target outside of the COSY tunnel, and after installation,  {\bf 2 weeks + 2 MD}  are needed, earliest in 2023.
		\item The first measurements with the new detector system will address the 		determination of the target polarization when H (or D) is injected into the 		storage cell. The task will be accomplished by making use of the single 		spin-asymmetries in $pd$ scattering. Scattered particles will be detected and 		identified by the new detector system surrounding the target cell. The comparison between the polarization seen by the beam as measured by the detector with the sampled polarization of the Breit-Rabi polarimeter will allow us to absolutely calibrate the polarimeter itself. 
		\item Mile stones: High target polarization and intensity, polarization
		measurement with the detector.
	\end{itemize}
	\item {\bf Longitudinal spin filtering} 
	\begin{itemize}
		\item After the success of steps 1. and 2., {\bf 5 weeks + 2 MD} are needed, earliest in 2023/24.
		\item The longitudinal spin filtering measurement will require the use of the Siberian snake to assure the longitudinal direction of the beam polarization at the polarized target place and of the new detection system to measure the induced polarization of the beam. A possible scheme of the measurement cycle will be:
		\begin{itemize}
			\item Beam injection and preparation.
			\item Switching on of the polarized internal target and start of the
			spin-filtering process. 
			\item At the end of the cycle, rapid inversion of the direction of the target
			polarization and measurement of the induced polarization in the beam.
			\item Alternatively, the beam polarization can be measured \textit{during} the
			polarization buildup by shortly switching the direction of the target
			polarization.
			\item Repetition of the measurement with opposite direction of the target
			polarization.
		\end{itemize}
	\end{itemize}
	\item {\bf Tests of new cell surfaces} 
	\begin{itemize}
		\item This development project for the use of storage cells at high-energy accelerators requires {\bf 2 weeks + 1 MD} in 2024 (see Sec.\,\ref{sec:cellcap} for more details).
		\item 	The existing technology of target cell coatings makes use of cell coatings 		like Dryfilm or Teflon to reduce recombination and depolarization on the walls of the storage cell of the polarized gas target. However, restriction rules at high energy accelerators don't allow the use of these materials inside the ring vacuum system. Therefore, new materials need to be tested for their behavior regarding recombination and preservation of polarization of the target gas. Tests in the laboratory have only limited validity, and therefore final tests in an accelerator under realistic conditions are necessary.
	\end{itemize}
\end{enumerate}

\subsubsection{Longitudinal polarization buildup}

A detailed discussion of the estimation of the longitudinal polarization buildup is provided in Appendix\,\ref{app:pax}. Figure\,\ref{fig:buildup-cs} shows the transverse and longitudinal polarization buildup cross sections $\tilde{\sigma}_1^{\rm T}$ and $\tilde{\sigma}_1^{\rm L}$. The blue region shows the  region where the beam can by cooled by electrons. For protons at $T = \SI{120}{MeV}$, the estimated longitudinal cross section is $\tilde{\sigma}_1^{\rm L} = \SI{-14.3}{mb}$, which is roughly half of the transverse cross section measured during the previous spin-filtering test at
COSY\,\cite{AUGUSTYNIAK201264}.

As discussed in Appendix\,\ref{app:pax}, with a target thickness of $d_{\rm t} = 5.50 \cdot 10^{13}\,{\rm atoms/cm^2}$, a COSY revolution frequency of $f = 753\,{\rm kHz}$, a target polarization of $Q = 0.8$, and a ring acceptance of $6.5\,{\rm  mrad}$, we obtain the longitudinal polarization buildup as function of time, shown in Fig.\,\ref{fig:buildup-2}. For a proton energy of $T = \SI{120}{MeV}$, the
estimated buildup rate is 0.0017/hour, which means that $\approx \SI{12}{\hour}$ of filtering time is required to reach a longitudinal polarization of 0.02.
\begin{figure}[tb]
	\subfigure[\label{fig:buildup-cs} Estimated longitudinal (black) and transverse (red) polarization buildup cross section as function of proton energy. The red triangle shows the transverse cross-section from the spin-filtering test at COSY\,\cite{AUGUSTYNIAK201264}.]{\includegraphics[height=0.24\textheight]{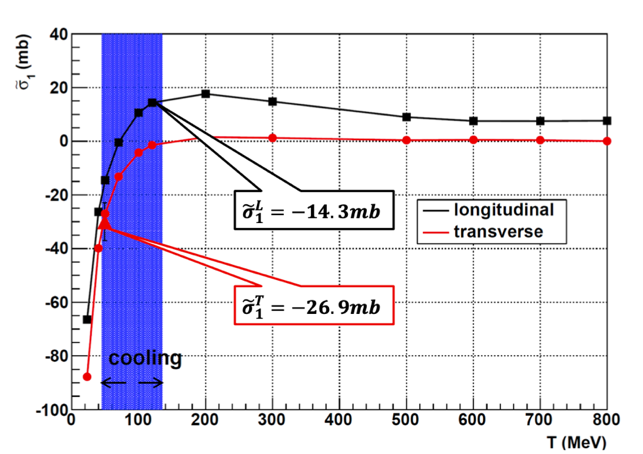}} \hspace{0.4cm}
	\subfigure[\label{fig:buildup-2} Estimated longitudinal polarization buildup as function of time, with the parameters described in the text. ]{\includegraphics[height=0.23\textheight]{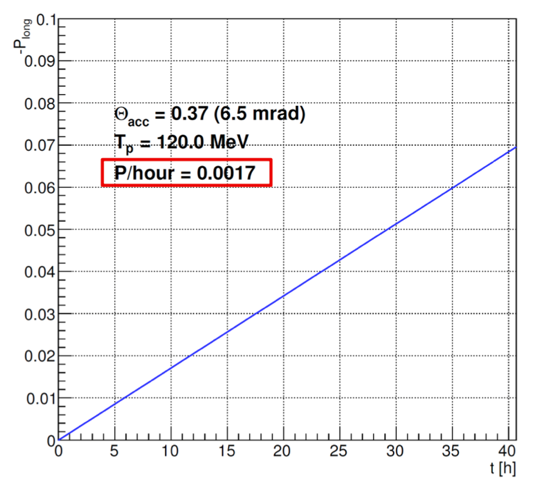}}
	\caption{(a) Estimated longitudinal spin-filtering cross section $\tilde{\sigma}^\text{L}_1$, and (b) polarization buildup at COSY as function of time for a proton beam kinetic energy of $T = \SI{120}{MeV}$.}
	\label{fig:buildup}
\end{figure}

\subsubsection{Tests of new target cell surfaces}
\label{sec:cellcap}
\begin{figure}[tb]
	\includegraphics[height=5.9cm]{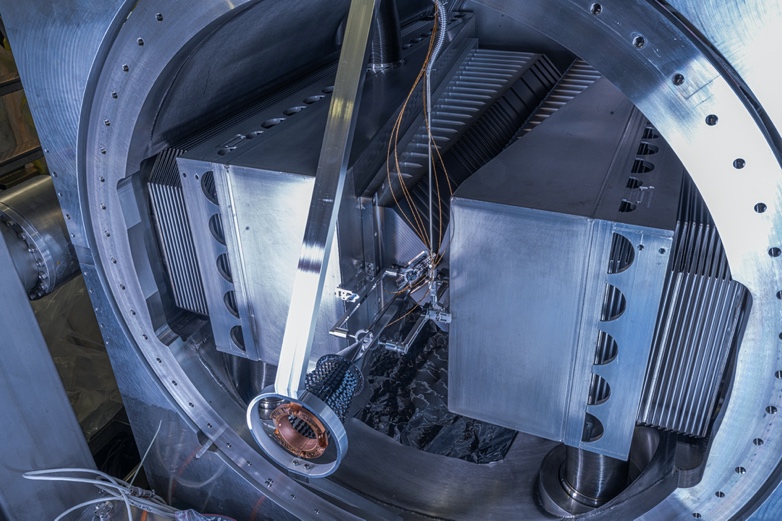}\hfill
	\includegraphics[height=5.9cm]{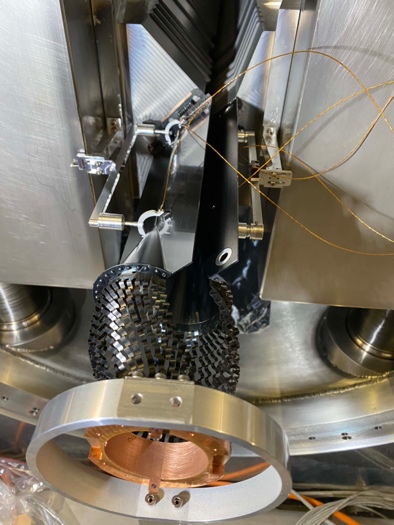}
	\caption{The openable target cell at the interaction point of LHCb.}
	\label{SMOG2}
\end{figure}

Over the last 20 years, the J\"ulich group together with its external collaborators has developed a solid and proven worldwide experience in the design and construction of storage cells for gaseous targets. The use of this technology has had a significant impact in the field of experimental hadronic physics. Examples of this application are the target of the HERMES experiment in HERA (DESY), used from 1995 to 2007, that of the OLYMPUS experiment in DORIS (DESY), used in the period 2012-2013, and that of the PAX  experiment, currently in use at COSY. The storage cell, typically consisting of a $50$ to $\SI{200}{\micro \meter}$ thick cylindrical aluminum tube, is placed inside the accelerator beam tube, coaxial to the beam. The acquired experience has recently allowed the development and construction of an openable storage cell for the LHCb experiment (see Fig.\,\ref{SMOG2} and SMOG2 project\,\cite{Barschel:2020drr,Aidala:2019pit}).
The storage cell has been successfully installed upstream of the LHCb detector during Summer  2020. Starting from RUN3 (2021), LHCb will therefore be the only LHC experiment to be
equipped with two distinct points of interaction,  and the ability to operate simultaneously in two collision modes: collider and fixed target. The collisions with the gas will occur with an energy in the center of mass of \SI{115}{GeV} for proton beams and \SI{72}{GeV} for lead beams. SMOG2 will make it possible to carry out precision studies in the field of QCD and  astroparticle physics in kinematic regions never probed before.

The next step of the development is related to the installation of a polarized atomic beam source. The performance of the storage cell for this case is strongly connected with the behavior of the cell surface with respect to recombination and depolarization of the stored polarized atomic hydrogen. Little is presently known about amorphous carbon, which is the only coating admitted for the LHC surfaces directly exposed to the beam. The PAX interaction point at COSY being equipped atomic beam source, Breit-Rabi polarimeter and vertex silicon detector is the only place worldwide where a characterization of the cell coating can be performed. Such a test can be conducted right after the completion of the longitudinal spin-filtering experiment.


\subsubsection{Appendix: Estimation of the longitudinal polarization buildup}	
\label{app:pax}

The rate of change of the number of particles $N$ and their total spin
$s=N_{\leftarrow}(t) - N_{\rightarrow}(t)$  is
described by:
\begin{equation}
\frac{d}{dt}\left(\begin{array}{c} N \\ s \end{array}\right) = -f
dt \begin{pmatrix} \tilde{\sigma}_0 & Q\cdot\tilde{\sigma}_1
\\ Q\cdot(\tilde{\sigma}_1+\Delta \sigma_1) &
\tilde{\sigma}_0+2\Delta\sigma_0 \end{pmatrix}\cdot \left(\begin{array}{c} N
\\ s \end{array}\right)
\end{equation}
where the contributing cross-sections are given by
\begin{eqnarray}
\tilde{\sigma}_0 &=& 2 \pi \int_{\theta_{\rm acc}}^{\theta_{\rm max}}
\frac{d\sigma}{d\Omega} \sin\theta d\theta \,,\\
\tilde{\sigma}_1^{\rm L} &=& 2 \pi \int_{\theta_{\rm acc}}^{\theta_{\rm max}} A_{00kk}
\frac{d\sigma}{d\Omega} \sin\theta d\theta \,,\\
\Delta\sigma_0^{\rm L} &=& 2 \pi \int_{\theta_{\rm min}}^{\theta_{\rm acc}}
(1+D_{s0k0}\sin\theta-D_{k0k0}\cos\theta)\frac{d\sigma}{d\Omega} \sin\theta
d\theta\,,\\
\Delta\sigma_1^{\rm L} &=& 2 \pi \int_{\theta_{\rm min}}^{\theta_{\rm acc}}
(A_{00kk}+K_{s00k}\sin\theta-K_{k00k}\cos\theta)\frac{d\sigma}{d\Omega}
\sin\theta d\theta\,.
\end{eqnarray}
Here $A$ is the double-spin asymmetry, $K$ indicates the polarization transfer
observables and $D$ the depolarization observables, as defined in
\cite{Bystricky:1976jr}. 

Figure\,\ref{fig:buildup-cs} shows the transverse and longitudinal polarization buildup cross sections $\tilde{\sigma}_1^{\rm T}$ and $\tilde{\sigma}_1^{\rm L}$. The blue region shows the  region where the beam can by cooled by electrons. For protons at $T = \SI{120}{MeV}$, the estimated longitudinal cross section is $\tilde{\sigma}_1^{\rm L} = \SI{-14.3}{mb}$, which is roughly half of the transverse cross section measured during the previous spin-filtering test at
COSY\,\cite{AUGUSTYNIAK201264}.

The buildup-rate has been determined by making use of the following coupled evolution equations for the polarization $P(t)$ and the beam intensity $I(t)$,
\begin{eqnarray}
P(t) &=& \frac{-Q(\tilde{\sigma}_1 +\Delta\sigma_1)\cdot\tanh(Q\sigma_3 d_{\rm t}
	ft)}{Q\sigma_3 + \Delta\sigma_0 \cdot Q \sigma_3 d_{\rm t} ft}\\
I(t) &=& I_0 \cdot e^{-(\tilde{\sigma}_0 + \Delta\sigma_0) d_{\rm t} f t} \cdot
\left[\cosh(Q\sigma_3 d_{\rm t} ft) + \frac{\Delta\sigma_0}{Q\sigma_3}
\sinh(Q\sigma_3 d_{\rm t} ft) \right]\,.
\end{eqnarray}
By using $Q\sigma_3 = \sqrt{Q^2\tilde{\sigma}_1(\tilde{\sigma}_1 + 	\Delta\sigma_1) + \Delta\sigma_0^2}$,  we obtain the following expressions,
\begin{eqnarray}
P(t) &\approx& -Q d_{\rm t} ft (\tilde{\sigma}_1 + \Delta\sigma_1)\\
I(t) &\approx& I_0 e^{-\frac{\displaystyle t}{\displaystyle \tau}}\,.
\end{eqnarray}
With a target thickness of $d_{\rm t} = 5.50 \cdot 10^{13}\,{\rm atoms/cm^2}$, a COSY revolution frequency of $f = 753\,{\rm kHz}$, a target polarization of $Q = 0.8$, and a ring acceptance of $6.5\,{\rm  mrad}$, we obtain the longitudinal polarization buildup as function of time, shown in Fig.\,\ref{fig:buildup-2}. For a proton energy of $T = \SI{120}{MeV}$, the
estimated buildup rate is 0.0017/hour, which means that $\approx \SI{12}{\hour}$ of filtering
time is required to reach a longitudinal polarization of 0.02.

\cleardoublepage

\section{Summary}
\subsection{Total beam time estimate}

Table\,\ref{tab:summary-table} summarizes the estimated number of machine development and beam time weeks in the period 2021 -- 2024 for the spin physics investigations presented in Sec.\,\ref{sec:spin-physics-at-COSY}.

\begin{table}[htb]
	\centering
	\begin{threeparttable}[b]
	\begin{tabular}{cp{7cm}cc}
		\hline
&		Experiment & Machine development & Beam time \\ \hline		
	Sec.\,\ref{sec:spico-for-protons} &	Spin-coherence time investigations for protons & \textcolor{red}{6} & \textcolor{red}{9} \\
Sec.\,\ref{sec:spintransparency} &	Spin transparency experiments for proton polarization control at integer spin resonances & \textcolor{red}{6} & \textcolor{red}{9} \\		
Sec.\,\ref{sec:spin-tune-orbit-bumps}	& Spin-tune mapping with orbit bumps & \textcolor{red}{4} & \textcolor{red}{8} \\	 
Sec.\,\ref{sec:axion}	& Axion search & \textcolor{red}{(8)} \tnote{1} & \textcolor{red}{(52)} \tnote{1} \\	
Sec.\,\ref{sec:cooling} & Interplay of spin-coherence time and stochastic cooling & \textcolor{red}{--} \tnote{2} & \textcolor{red}{--} \tnote{2} \\
Sec.\,\ref{sec:spinfiltering} & Longitudinal spin filtering with protons & \textcolor{red}{7} & \textcolor{red}{11} \\\hline 
& \hspace{6cm} Total & \textcolor{red}{23}& \textcolor{red}{37}\\\hline
		\end{tabular} 
\caption{\label{tab:summary-table} Estimated machine development and beam time (in weeks) for the indicated experimental investigations using COSY. }
		\begin{tablenotes}
			\item[1] \footnotesize Not included in the total numbers (for further details see footnote \ref{footnote:2.4} in Sec.\,\ref{sec:axion}).
			\item[2] \footnotesize Project not yet in a state that allows estimates. 
		\end{tablenotes}
	\end{threeparttable}
\end{table}

\subsection{Perspectives for the first direct measurement of the proton electric dipole moment}
\label{sec:protonedm}

The first deuteron EDM measurement have recently been pursued by the JEDI collaboration\,\cite{Rathmann:2013rqa,PhysRevSTAB.16.114001}. The primary reason to begin storage ring EDM studies with deuterons is the absence of depolarizing resonances that might adversely affect the spin-coherence time. 

Once a sufficiently long proton spin-coherence time ($\approx \SI{1000}{s}$) has been achieved in conjunction with a thorough understanding of the connection between spin-coherence, chromaticity and beam emittance growth, a first direct proton EDM measurement can be also performed at COSY. The RF Wien filter technique can be applied to the stored longitudinally polarized protons to accumulate the EDM signal, as described, e.g., in Ref.\,\cite{PhysRevAccelBeams.23.024601}. Recently, the technique has been improved by the  application of a  spin phase-lock feedback system acting on individual bunches stored in the machine\,\cite{proposal-Jamal-pilot-bunch}. Although, due to its larger magnetic dipole moment, the spin-precession frequencies for protons are substantially larger than for deuterons, no principal difficulties are anticipated. The phase-lock technique, pioneered by JEDI, will be applied as well\,\cite{PhysRevLett.119.014801}. Recently, the technique has been successfully improved by the  application of a spin phase-lock feedback acting on individual bunches stored in the machine\,\cite{proposal-Jamal-pilot-bunch}.

The measurement will represent the first direct measurement of the proton EDM and pave the way to the next stage of the research, namely the design and construction of a dedicated prototype ring.


\clearpage
 
%

\bibliographystyle{apsrev-title}
\bibliography{SPAC}

\end{document}